\documentclass[fleqn,usenatbib]{mnras}

\usepackage{newtxtext,newtxmath}
\usepackage[T1]{fontenc}

\DeclareRobustCommand{\VAN}[3]{#2}
\let\VANthebibliography\thebibliography
\def\thebibliography{\DeclareRobustCommand{\VAN}[3]{##3}\VANthebibliography}

\usepackage{graphicx}	% Including figure files
\newcommand{\Ro}{\ensuremath{{R_{\mathrm o}}}}
\newcommand{\Rfir}{\ensuremath{R_{\rm f/i}}}
\newcommand{\taustar}{\ensuremath{\tau_{\ast}}}

\newcommand{\xspec}{\textsc{xspec}}
\newcommand{\apec}{\textsc{apec}}
\newcommand{\atomdb}{\textsc{atomdb}}
\newcommand{\tlusty}{\textsc{tlusty}}
\newcommand{\vinf}{\ensuremath{v_{\infty}}}

\newcommand{\chandra}{\textit{Chandra}}
\newcommand{\xmm}{\textit{XMM}}
\newcommand{\Rstar}{\ensuremath{{R_{\ast}}}}

\newcommand{\zpup}{$\zeta$ Pup}

\newcommand{\hetgs}{HETGS}
\newcommand{\meg}{MEG}
\newcommand{\heg}{HEG}
\newcommand{\kms}{km s$^{-1}$}
\newcommand{\Teff}{\ensuremath{T_{\rm eff}}}

\newcommand{\beq}{\begin{equation}}
\newcommand{\eeq}{\end{equation}}
\newcommand{\beqa}{\begin{eqnarray}}
\newcommand{\eeqa}{\end{eqnarray}}

\DeclareGraphicsExtensions{.pdf,.png,.jpg}

\title[\zpup\/ \chandra\/ He-like complexes]{Helium-like X-ray line complexes show that the hottest plasma on the  O supergiant $\zeta$ Puppis is in its wind}

\author[D. H. Cohen et al.]{David H.\ Cohen,$^{1}$\thanks{E-mail:
    dcohen1@swarthmore.edu} 
Ariel M.\ Overdorff,$^{1}$ 
Maurice A.\ Leutenegger,$^{2}$ 
\newauthor Marc Gagn\'e,$^{3}$ 
V\'{e}ronique Petit,$^{4}$ 
Alexandre David-Uraz$^{5,2,6}$ 
\\
$^{1}$Swarthmore College, Department of Physics and Astronomy, Swarthmore, Pennsylvania 19081, USA\\
$^{2}$NASA/Goddard Space Flight Center, Code 662, Greenbelt, Maryland 20771, USA \\
$^{3}$West Chester University, Department of Geology and Astronomy, West Chester, Pennsylvania 19383, USA \\
$^{4}$University of Delaware, Department of Physics and Astronomy, Newark, Delaware 19716, USA \\
$^{5}$ Department of Physics and Astronomy, Howard University, Washington, DC 20059, USA \\
$^{6}$ Center for Research and Exploration in Space Science and Technology, NASA/GSFC, Greenbelt, MD 20771, USA
}

\date{Accepted 28 March 2022. Received 24 March 2022; in original form 4 October 2021}

\pubyear{2022}

\begin{document}
\label{firstpage}
\pagerange{\pageref{firstpage}--\pageref{lastpage}}
\maketitle

\begin{abstract}
We present an analysis of \chandra\/ grating spectra of key helium-like line complexes to put constraints on the location with respect to the photosphere of the hottest ($T \gtrsim{6 \times 10^6}$ K) plasma in the wind of the O supergiant \zpup\/ and to explore changes in the 18 years between two sets of observations of this star. We fit two models -- one empirical and one wind-shock-based -- to the S {\sc xv}, Si {\sc xiii}, and Mg {\sc xi} line complexes and show that an origin in the wind flow, above $r \approx 1.5$ \Rstar, is strongly favored over an origin less than 0.3 \Rstar\/ above the photosphere ($r \lesssim 1.3$ \Rstar), especially in the more recent, very long-exposure data set. There is a modest increase in the line and continuum fluxes, line widths, wind absorption signatures, and of the hot plasma's distance from the photosphere in the 18 years since the first \chandra\/ grating observation of \zpup. Both modes of modeling include the effects of dielectronic recombination satellite emission line blending on the helium-like complexes -- the first time this has been accounted for in the analysis of He-like line ratios in O stars. 
\end{abstract}


\section{Introduction}

X-rays from single O stars are generally thought to arise in these stars' winds, where the line-deshadowing instability (LDI) \citep{ocr1988} produces shock-heated plasma distributed in numerous locations above an onset radius that theory and observations both suggest is $R_{\rm o} \approx 1.5$ \Rstar\/ \citep{fpp1997,Kahn2001,ro2002,Cohen2014b}. There are however persistent suggestions -- based on interpretations of key line ratios in helium-like lines of metals --  that the hottest X-ray plasma may arise much closer to the photosphere, where the LDI wind-shock mechanism should not be able to generate plasma of the observed temperatures ($T \gtrsim{10^7}$ K) and magnetic heating mechanisms are then invoked to explain the X-ray observations \citep{wc2001,Cassinelli2001,wc2007}. Additionally, recent multi-dimensional hydrodynamical simulations of the LDI show shock onset radii of less than $1.5$ \Rstar, although it is not clear that the shocks in this region are strong enough to produce X-rays \citep{Sundqvist2018}. In the LDI context, the He-like diagnostics can provide constraints on the strong-shock-onset radius.

Recent \chandra\/ High Energy Transmission Grating Spectrometer (\hetgs) spectra of \zpup\/ (O4If) present an opportunity to make the most precise measurements yet in any single non-magnetic O star of the key diagnostic lines in the complex of transitions from $n = 2$ to the ground state in helium-like ions. This complex contains, from shortest to longest wavelength, the resonance line ($^1{\rm P_1} \rightarrow$ $^1{\rm S_0}$), the intercombination line ($^3{\rm P_{2,1}} \rightarrow$ $^1{\rm S_0}$), and the forbidden line ($^3{\rm S_1} \rightarrow$ $^1{\rm S_0}$) \citep{gj1969}. The wavelength separation of these three\footnote{The intercombination ``line'' is actually two very closely spaced lines which, with the Doppler broadening due to the wind's high velocity, functionally constitute a single line feature.} lines is roughly equal to the Doppler-broadened line widths in the $\vinf = 2250$ \kms\/ \citep{Haser1995} wind of \zpup. The lines therefore are blended and cannot reliably be measured separately but rather need to be fit simultaneously with a model. 

The key diagnostic of the hot plasma location is the ratio of the strength of the forbidden line to that of the intercombination line. In coronal equilibrium (steady-state level populations governed by collisional excitation from the ground state and spontaneous decay from excited levels) the forbidden-to-intercombination line ratio ($\mathcal{R} \equiv f/i$) is roughly 3 (with variation from element to element). However, photoexcitation (or at high enough densities, collisional excitation) can move electrons from the metastable upper level of the forbidden transition to the upper level(s) of the intercombination line, strengthening the latter at the expense of the former. Thus, the measured $f/i$ ratio is sensitive to the local ultraviolet (UV) radiation field at the wavelength of the key transition between the two excited states \citep{gj1969,bdt1972,Porquet2001}. With an estimate of the photospheric flux at the relevant UV wavelengths, the $f/i$ line ratio becomes a diagnostic of the distance of the X-ray emitting plasma from the photosphere \citep{Porquet2001,Kahn2001,wc2007}. In the case that embedded wind shocks explain a star's X-ray emission, the diagnostic tells us about the characteristic location of the shock-heated plasma or, in the context of a model of the spatial distribution of hot plasma, the onset radius of shock emission and the assumed filling factor above that \citep{Leutenegger2006}.

Because of the dependence on atomic number of the relevant transition rates and of the relevant UV wavelengths for photoexcitation, each helium-like ion has a different sensitivity to distances of interest, with higher-Z ions more effectively probing the base of the wind, and lower-Z ions the outer parts of the wind. For \zpup, which has an effective temperature of 40,000 K \citep{Howarth2019}, the most sensitive line complexes for studying the base of the wind are those of S {\sc xv}, Si {\sc xiii}, and Mg {\sc xi}, near 5.0, 6.7, and 9.2 \AA, respectively. These ions, especially S {\sc xv}, reside in the hottest plasma producing observable X-rays in \zpup. The temperatures of peak line emissivity are $1.3$ to $1.6 \times 10^7$ K, $10^7$ K, and $6 \times 10^6$ K for S {\sc xv}, Si {\sc xiii}, and Mg {\sc xi}, respectively \citep{Bryans2006, Bryans2009, Foster2012}.  In the cycle 1 \chandra\/ observation of \zpup, obtained in 2000, the S {\sc xv} line complex is the one that indicates a formation radius closest to the photosphere: $R_{\rm f/i} < 1.2$ \Rstar\/ \citep{Cassinelli2001}, $R_{\rm f/i} = 1.1^{+0.4}_{-0.1}$ \Rstar\/ \citep{Leutenegger2006}, $R_{\rm f/i} < 1.22$ \Rstar\/ \citep{wc2007}\footnote{When denoting a single radius of X-ray formation based on measured forbidden-to-intercombination line flux ratios, we use the notation $R_{\rm f/i}$.}. 

The cycle 19 \chandra\/ observations, obtained in 2018-19, have a much longer aggregate exposure time than the single observation taken in cycle 1. The better signal-to-noise of the more recent data allows us to place tighter constraints on the X-ray plasma location than those obtained from the older, shorter exposure time data. They also enable us to look for changes in the X-ray formation location in the 18 years between the observations. The hot plasma emission measure \citep{Huenemoerder2020} and emission line fluxes \citep{Cohen2020} have increased since the earlier observations, and the line profile shapes have also changed in a way consistent with a mass-loss rate increase \citep{Cohen2020}, so other changes to the wind X-ray properties may also be evident. 

The subject of this study, \zpup\/ \citep[HD 66811, O4 I (n)fp;][]{Sota2014}, is the brightest early O star in the sky and has long been a prototype for studying the atmospheres and winds of O stars. A recent detailed analysis of the {\it Hipparcos} data by \citet{Howarth2019} confirms the parallax of $3.01\pm0.10$ mas and the resulting distance of $332\pm11$~pc. Unfortunately, at a visual magnitude of 2, $\zeta$~Pup is too bright for inclusion in the {\it GAIA} EDR3/DR3 catalog \citep{GAIA2021}. This distance result is controversial because it is approximately half the distance inferred from detailed spectroscopic modeling \citep[e.g.,][]{Pauldrach2012}, implying that \zpup\/ is smaller, less luminous, and less massive than previously assumed. \citet{Howarth2019} suggest that the properties can be understood if the star is a merger product. In this paper we adopt the stellar parameters those authors determined using the distance of 332 pc: $T_{\rm eff} = 40,000$~K, $\log g = 3.58$,  $\log (L_{\ast}/L_{\odot}) = 5.65$, $R_{\ast} = 13.5 ~{\rm R_{\odot}}$, and $M_{\ast} = 20 ~{\rm M_{\odot}}$. These are directly relevant to the work discussed in this paper only in terms of their impact on the UV photospheric flux from the stellar atmosphere model choice discussed in the next section. 

Low-amplitude periodic and quasi-periodic variability in the optical \citep[summarized in][]{Howarth2019} and UV \citep{Massa1995} is seen in \zpup. Notably, there is, at least in recent years, a persistent $1.78$-day periodicity in {\it BRITE} satellite optical photometry \citep{Tahina2018} that is also seen in these same cycle 19 {\it Chandra} data \citep{Nichols2021}. The 1.78 day periodicity in the \chandra\/ data has a peak-to-peak amplitude of 6 percent, and there is also a random component to the variability with a root-mean-square variation of 4.5 percent \citep{Nichols2021}. 

In \S2 we describe the data and in \S3 we describe the models and the fitting procedure. In \S4 we report the results and in \S5 we discuss the implications for models of X-ray production in O stars. Modeling of dielectronic recombination satellite lines that are blended with the He-like lines is discussed in an appendix. 

\section{The data}

More than 813 ks of pointed \chandra\/ observations of \zpup, comprising 21 separate observations, were made during cycle 19 (2018-19, PI: Waldron). We downloaded all the data from the archive, along with the single, 68 ks cycle 1 observation from 2000, and reprocessed them using {\sc caldb} v.\ 4.8.5. We extracted the first order spectra from each on-axis pointing using standard {\sc ciao} tools, and produced corresponding response matrix and grating auxiliary response (effective area) files. These data are discussed elsewhere \citep{Huenemoerder2020,Cohen2020}, and specifically, they show only modest variability from pointing to pointing, with low-amplitude variability, described in the previous section, and no long-term trends evident \citep{Nichols2021}. Furthermore, the X-ray line flux and emission measure changes between cycle 1 and cycle 19 are significantly bigger than the variation within cycle 19. We therefore analyze all the cycle 19 data simultaneously and given the long total exposure time, these data have roughly three times better signal-to-noise than the cycle 1 observation that was the basis for the previous helium-like $f/i$ analyses of this star. An observing log is available in table~1 of \citet{Cohen2020}. 

\section{Modeling the line complexes}

\noindent 
The $f/i$ line ratio diagnostic depends on the relative rates of photoexcitation and spontaneous emission out of the metastable excited state of the forbidden line and so is a function of atomic parameters and the local UV mean intensity. The wavelengths for the UV photoexcitation are shortward of the Lyman edge for S {\sc xv} and Si {\sc xiii} and so the photospheric fluxes are not directly observed and instead have to be estimated from models. 

The formalism for the line ratio's dependence on atomic parameters and on the UV local mean intensity is described in \citet{Leutenegger2006} and the relevant atomic parameters are collected there. From that paper, we have the following equations:

\begin{equation}
    \mathcal{R}(r) = \mathcal{R}_0 \frac{1}{1+2PW(r)},
    \label{eq:Rofr}
\end{equation}

\vspace{0.1in}
\noindent 
where $\mathcal{R}_0$ is the $f/i$ ratio in the absence of UV excitation. $P = \phi_*/\phi_c$, where $\phi_c$ scales with the spontaneous emission rate and $\phi_*$ is the photoexcitation rate $2^3S_1 \rightarrow 2^3P_J$ at the photosphere:

\begin{equation}
    \phi_* = 8 \pi \frac{\pi e^2}{m_e c} f \frac{H_\nu}{h \nu},
    \label{eq:phi}
\end{equation}

\vspace{0.1in}
\noindent 
in which $H_\nu$ is the surface Eddington flux at the excitation wavelength, $h\nu$ is the UV photon's energy, $\frac{\pi e^2}{m_{\rm e}c}$ is the classical bound electron cross section, and $f$ is the sum of the oscillator strengths of the transitions $2^3S_1 \rightarrow 2^3P_J$. Values for $f$ and $\phi_{\rm c}$ as well as the UV wavelengths of the relevant transitions, are taken from table 1 in \citet{Leutenegger2006}. Values of $\mathcal{R}_0 = 1.71$, $2.38$, and $2.91$, for S {\sc xv}, Si {\sc xiii}, and Mg {\sc xi}, respectively, are computed using the {\it Astrophysical Plasma Emission Code} \cite[\apec, ][]{Foster2012} -- see Appendix \ref{appendix:satellites} for details. Lastly, $W(r)$ is the geometrical dilution: 

\begin{equation}
    W(r) = \frac{1}{2} (1-(1-(R_*/r)^2)^{1/2}),
    \label{eq:W}
\end{equation}

\vspace{0.1in}
\noindent 
where $R_*$ is the star's radius and $r$ is the radial coordinate of the X-ray emitting plasma. This treatment does not account for the possible absorption or scattering of photospheric UV radiation as it propagates to the X-ray emitting plasma in the wind. 

The $f/i$ value for a particular element near the photosphere of a particular star is therefore uniquely determined by the surface flux and the distance of the X-ray emitting plasma from the photosphere. One traditional approach for determining the X-ray emitting plasma location is to measure the $f/i$ line flux ratio and then to model the measured value to constrain the location, $R_{\rm f/i}$. Another, more sophisticated but more model-dependent, approach is to assume a spatial distribution of the X-ray emitting plasma and compute the line fluxes at each location in order to synthesize a spectral feature \citep{Leutenegger2006}. The parameters of the model, including those that describe the spatial distribution, can then be fit directly. We use both approaches here, implemented in \xspec\/ \citep{Dorman2001} as {\it hegauss} and {\it hewind}, respectively\footnote{ See  \url{https://heasarc.gsfc.nasa.gov/xanadu/xspec/models/windprof.html}. 
}. We describe each model later in this section.

Before fitting either of these models, we fit the continuum emission with a power-law model with index $n = 2$ (which corresponds to a flat spectrum in wavelength) over the spectral regions on either side of the complexes. In the cases of S {\sc xv} and Mg {\sc xi} we omitted wavelength ranges with potentially-contaminating weak lines. These continuum fits are shown in Appendix A of \citet{Cohen2020}. Then the complexes themselves were fit with the relevant line complex model on top of this previously determined power-law continuum. The fitting was done by C statistic minimization \citep{Cash1979}, and we determined 68\% confidence limits using the ${\Delta}C$ formalism \citep{Nousek1989}. We fit the two spectral orders ($-1$ and $+1$) and data from the two grating arrays (MEG and HEG), and in the case of the cycle 19 data, the separate observations, simultaneously but without co-adding any of them.  

In addition to the weak continuum emission under the He-like lines, there are numerous, mostly weak, satellite lines generated by dielectronic recombination (DR) of He-like ions into the Li-like charge state, followed by cascades to the ground state in the presence of a spectator electron, producing lines with wavelengths similar to but generally a little bit longward of the corresponding He-like transitions. Traditionally in the type of analysis we present here, the satellite lines have been ignored, perhaps assumed to be approximately accounted for by the continuum modeling. The comprehensive paper by \citet{Porquet2001} does include a treatment of satellite line blending for Mg {\sc xi} and Si {\sc xiii}, assuming intrinsically narrow lines and instrumental broadening for several different grating spectrometers and several different temperatures and dilution factors. The application of the \citet{Porquet2001} models to wind Doppler-broadened lines in O star spectra is somewhat uncertain, however, due both to the large Doppler broadening seen in O star X-ray spectra and to the particular temperature distribution in the X-ray emitting plasma of these stars. 

Atomic models of DR satellites are complex, given the contributions from many different excited states, and early in the \chandra\/ and \xmm\/ era they were not readily available for data fitting. Currently, the commonly used \apec/\atomdb\/ spectral modeling code and database does include many DR satellite lines \citep{Foster2012,Foster2020}. In this paper we account for the satellite contributions in both the {\it hegauss} and {\it hewind} modeling by including the 29 (for S {\sc xv}) to 39 (for Mg {\sc xi}) strongest satellite lines in the data fitting in \xspec. We provide details of the satellite contribution and modeling and its effect on the derived He-like quantities in Appendix \ref{appendix:satellites}. All the results presented in the paper include the effects of this satellite contamination on the derived He-like line complex properties. 

The {\it hegauss} model assumes that each He-like complex is the sum of three Gaussian line profiles with identical widths and shifts but different line fluxes. The free parameters are the $f/i$ line flux ratio $\mathcal{R}$, the line flux ratio $\mathcal{G} = (f+i)/r$, and an overall normalization (flux) parameter, as well as a single line width, $\sigma_v$, and line shift, $\delta_v$, parameter. This parameterization avoids having to deal with covariance terms for fluxes from individual blended lines in the complex, and allows direct probing of the confidence interval for $\mathcal{R}$, the parameter of interest. After the $\mathcal{R}$ values are derived from this {\it hegauss} fitting, we model their radial dependence outside of \xspec. 

The {\it hewind} model is the sum of three {\it windprof} components \citep{OC2001, Leutenegger2006} that model the Doppler broadening according to an assumed wind beta-velocity law and partially optically thick transport through the wind, which naturally results in a blue-shifted and skewed line profile shape. At each radius, the relative contributions of the forbidden and intercombination lines are computed using the surface UV flux and atomic parameter values (see equations \ref{eq:Rofr} and \ref{eq:phi}). The free parameters of this model are $\tau_*$ -- a characteristic optical depth -- and the onset radius of X-ray emission, $R_0$, as well as the overall line complex flux. The $\mathcal{G}$ line ratio parameter described above for {\it hegauss}, which is mildly sensitive to the plasma temperature \cite[e.g.,][]{Porquet2001}, is also a free parameter of this model. We stress that in this {\it hewind} model, the forbidden-to-intercombination line flux ratio is not fit directly, but rather is computed at each radius and the information the model provides about the radial dependence is contained in the onset radius fit parameter, \Ro. Above \Ro, the X-ray emission is assumed to scale with the wind density squared, and the relative intensities of the forbidden and intercombination lines are controlled by the UV mean intensity at each radius \citep{Leutenegger2006}. 

\begin{figure}
\centering
    \includegraphics[angle=0,width=0.53\textwidth]{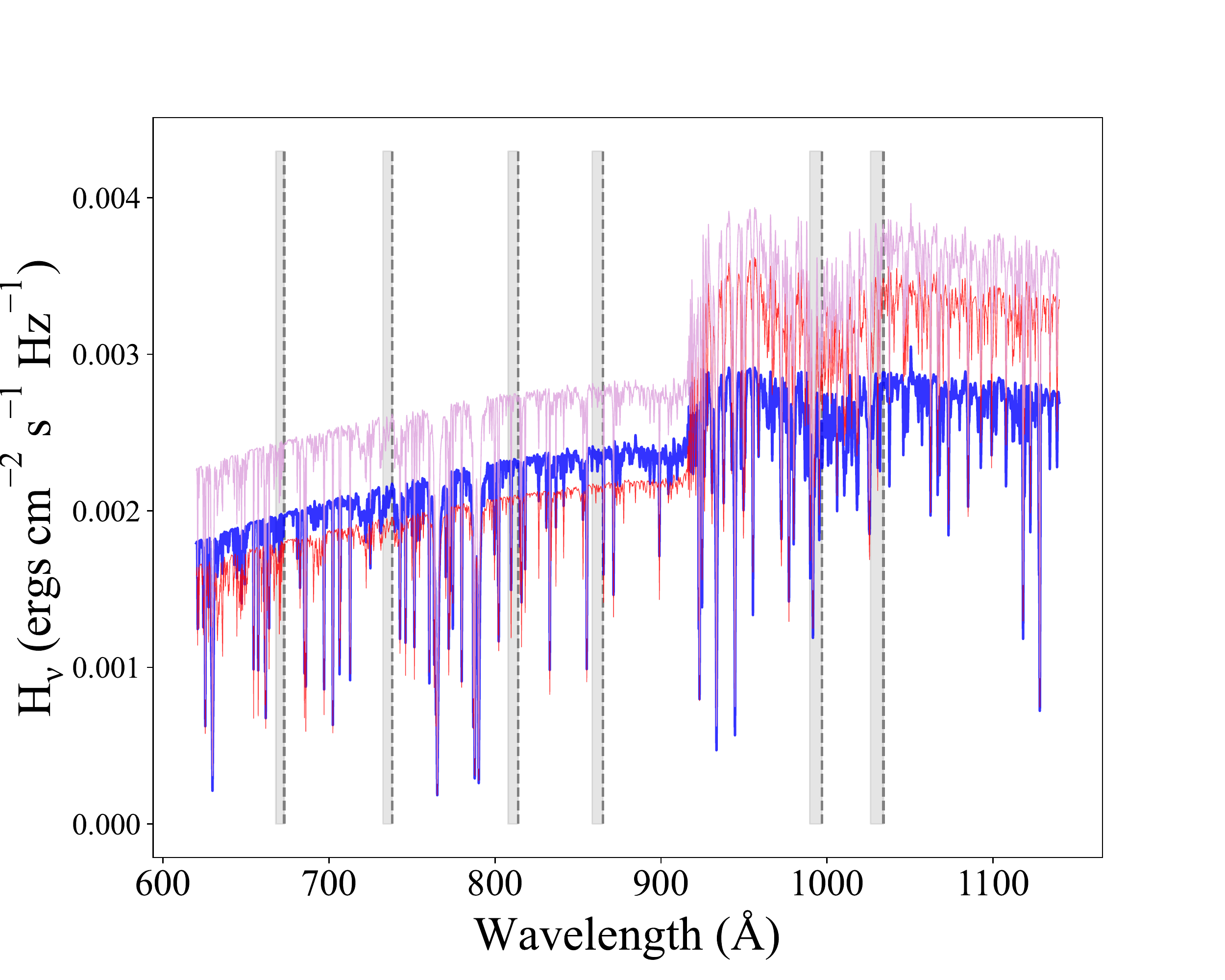}
    \caption{The \tlusty\/ model we adopted ($\Teff = 40,000$ K and $\log g = 3.50$) is shown in blue alongside two other models with slightly different parameter values ($\Teff = 40,000$ K and $\log g = 3.75$, in red; $\Teff = 42,500$ K and $\log g = 3.75$, in plum).  The vertical dotted lines indicate the driving wavelengths for the UV transitions in the three ions — the two shortest for S {\sc xv} to the two longest for Mg {\sc xi}, with the two in the middle at the wavelengths relevant for Si {\sc xiii}. The shaded region adjacent to each vertical line is the portion of the spectrum over which we averaged the flux to account for the wind Doppler shift. 
    }
	\label{fig:Gvs}
\end{figure}

We used \tlusty\/ model atmospheres \citep{Lanz2003} to obtain a value for the UV flux at the driving wavelengths for the transitions relevant to the $f/i$ ratio. To get a sense for the uncertainty on these flux values, we compared \tlusty\/ and \textsc{p}o\textsc{wr} atmosphere models \citep{Hainich2019}, which agreed well. We adopted the \tlusty\/ model with parameters closest to \zpup's -- $\Teff = 40,000$ K, $\log g = 3.50$. We show this model, along with the two models nearest it in the \tlusty\/ grid, in Fig.\ \ref{fig:Gvs}. Obvious limitations of this model for our use here include the fact that it has a low-rotational velocity and is spherically symmetric, while \zpup\/ has a projected rotational velocity of $v_{\rm rot} \sin i = 210$ km s$^{-1}$ and is likely at least modestly non-spherical \citep[for a comprehensive assessment, see][]{Howarth2019}, and also that these models assume Solar abundance while \zpup\/ has significant CNO processing evident in its spectrum \citep{Bouret2012,Martins2015}. 

In order to obtain a single flux value, $H_{\nu}$, for each ion to use in the X-ray modeling, we averaged the model atmosphere flux over a wavelength region corresponding to the blue-shifted wind Doppler broadening (shown graphically in Fig.\ \ref{fig:Gvs}). We note that this averages over quite a few photospheric absorption lines in each spectral region. We combine these average fluxes for the two transitions of a given ion -- S {\sc xv}: 0.00188, 0.00206; Si {\sc xiii}: 0.00224, 0.00234; Mg {\sc xi}: 0.00225, 0.00261 (ergs cm$^{-2}$ s$^{-1}$ Hz$^{-1}$) -- into a single average flux value for each ion (with $\frac{H_{\rm \nu}}{h\nu}$ weighted by the relevant transition probabilities) and use that in equation \ref{eq:phi}. We note that for the {\it hegauss} fitting, these UV fluxes are used in the subsequent modeling stage, where the $f/i$ line flux ratio derived from {\it hegauss} is modeled using equations \ref{eq:Rofr} through \ref{eq:W}. For the {\it hewind} fitting, $P$ in equation \ref{eq:Rofr} is specified as a parameter of the model in \xspec. The values we use are $P = 2.67, 15.7$, and $118.7$ for  S {\sc xv}, Si {\sc xiii}, and Mg {\sc xi}, respectively. 

\section{Results}

\begin{figure*}
\centering
    \includegraphics[angle=0,width=0.47\textwidth]{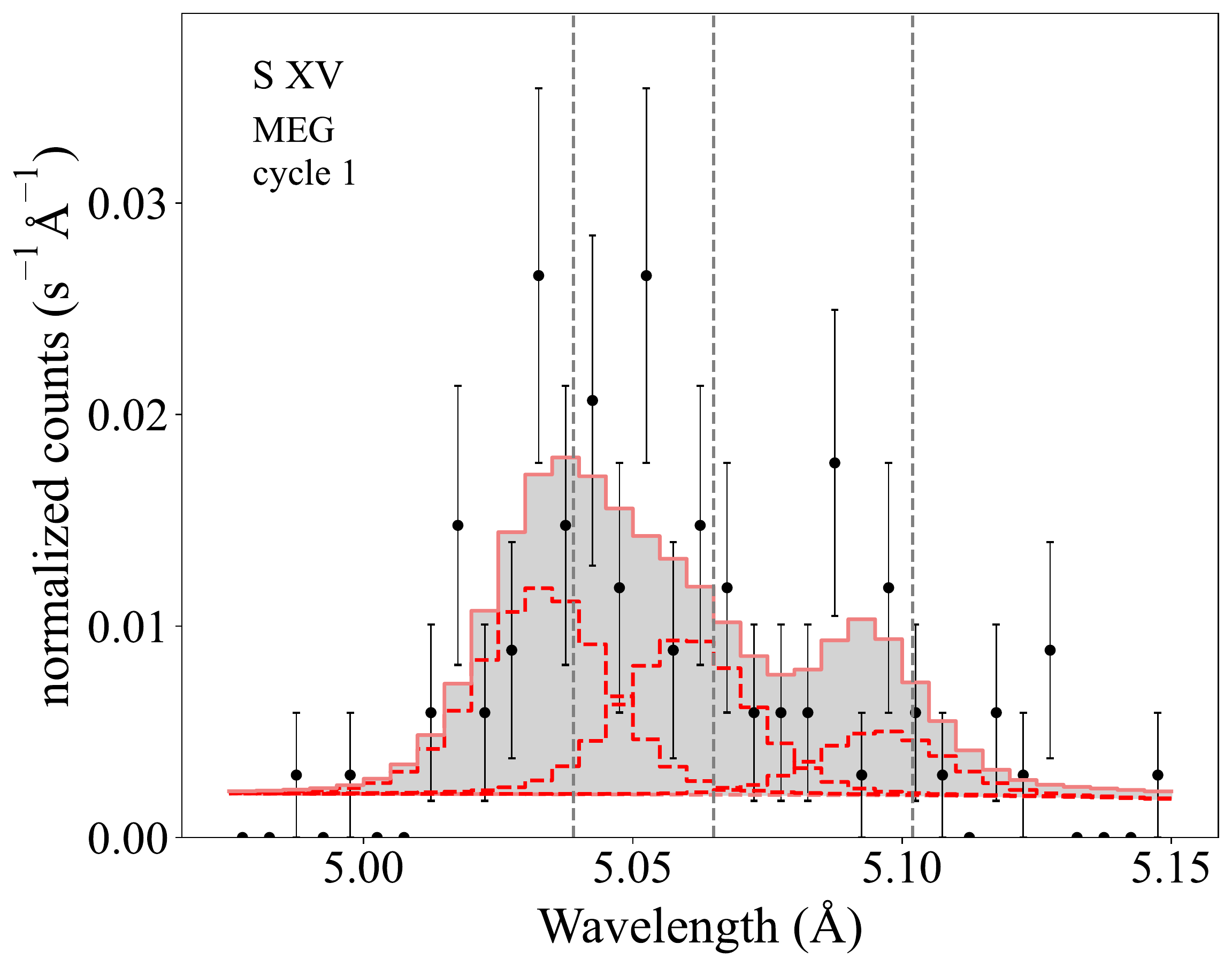}
    \includegraphics[angle=0,width=0.47\textwidth]{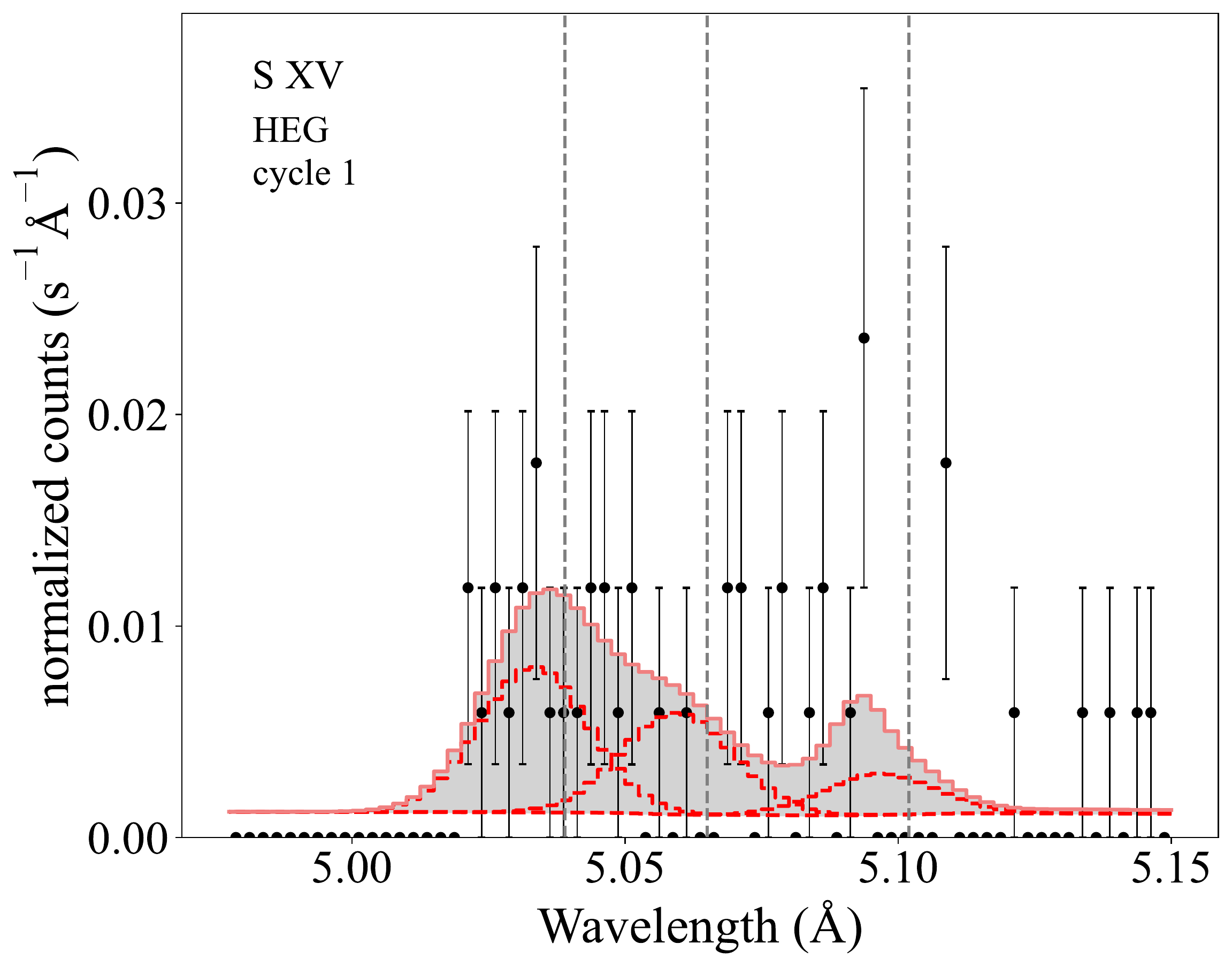}
    \includegraphics[angle=0,width=0.47\textwidth]{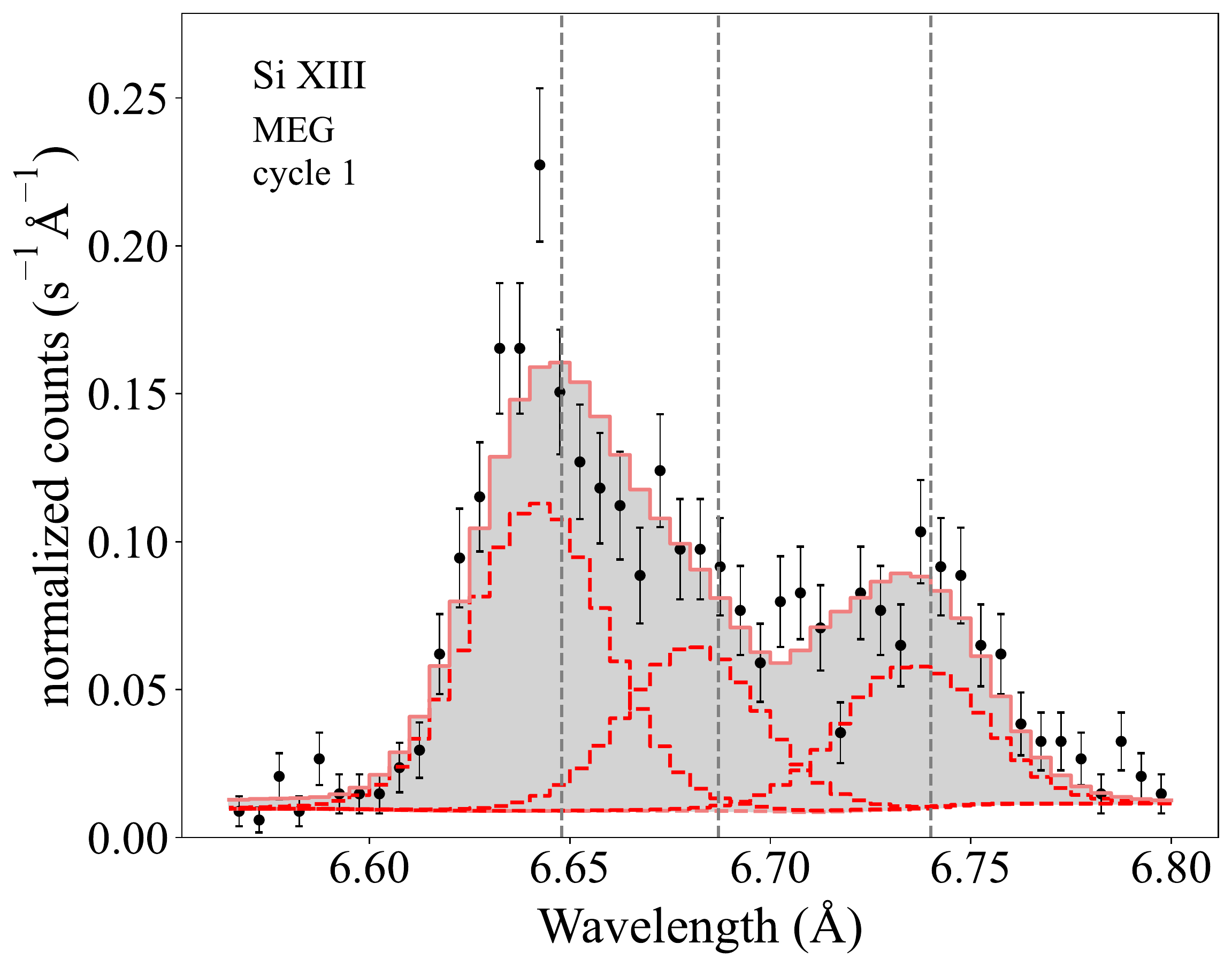}
    \includegraphics[angle=0,width=0.47\textwidth]{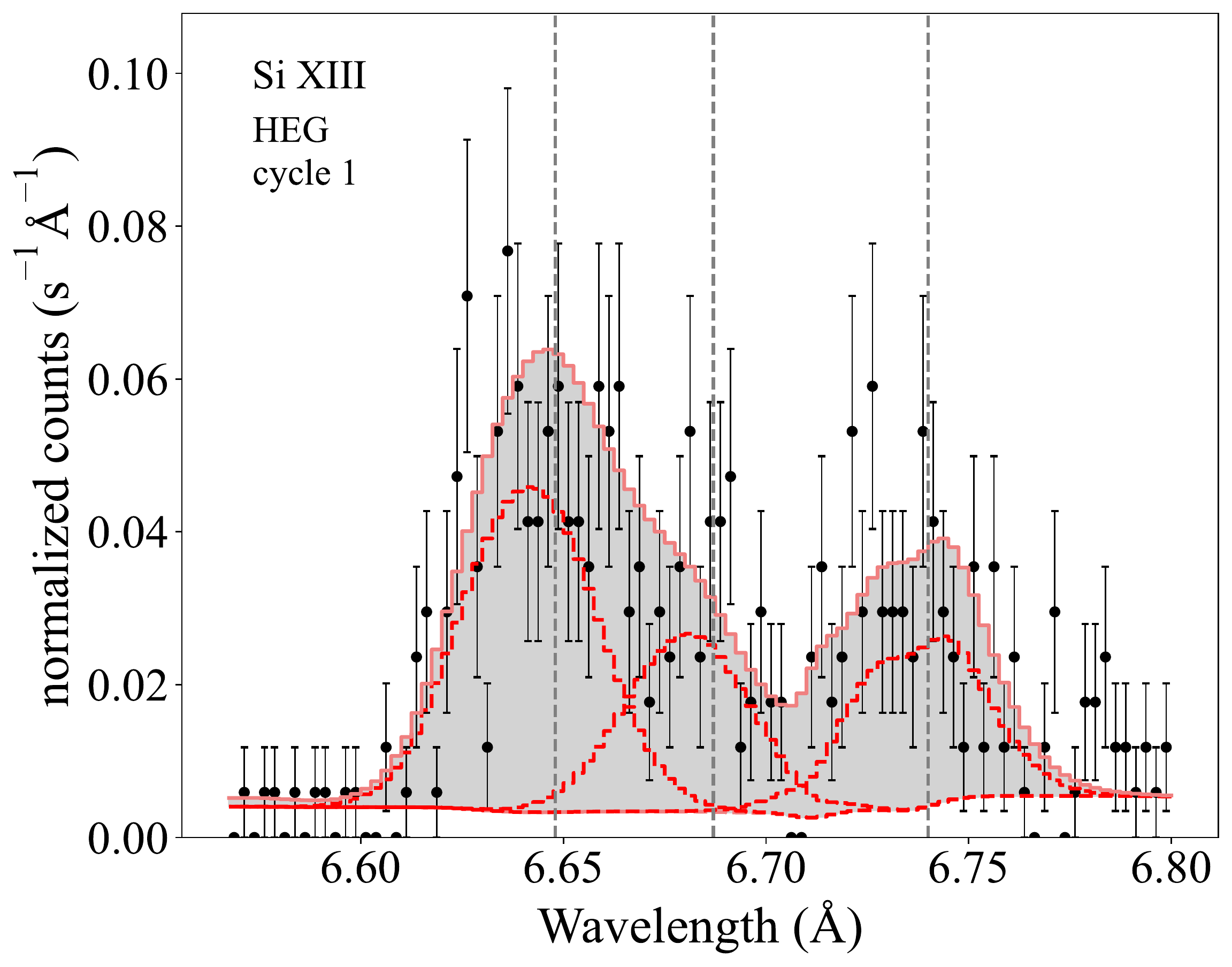}
    \includegraphics[angle=0,width=0.47\textwidth]{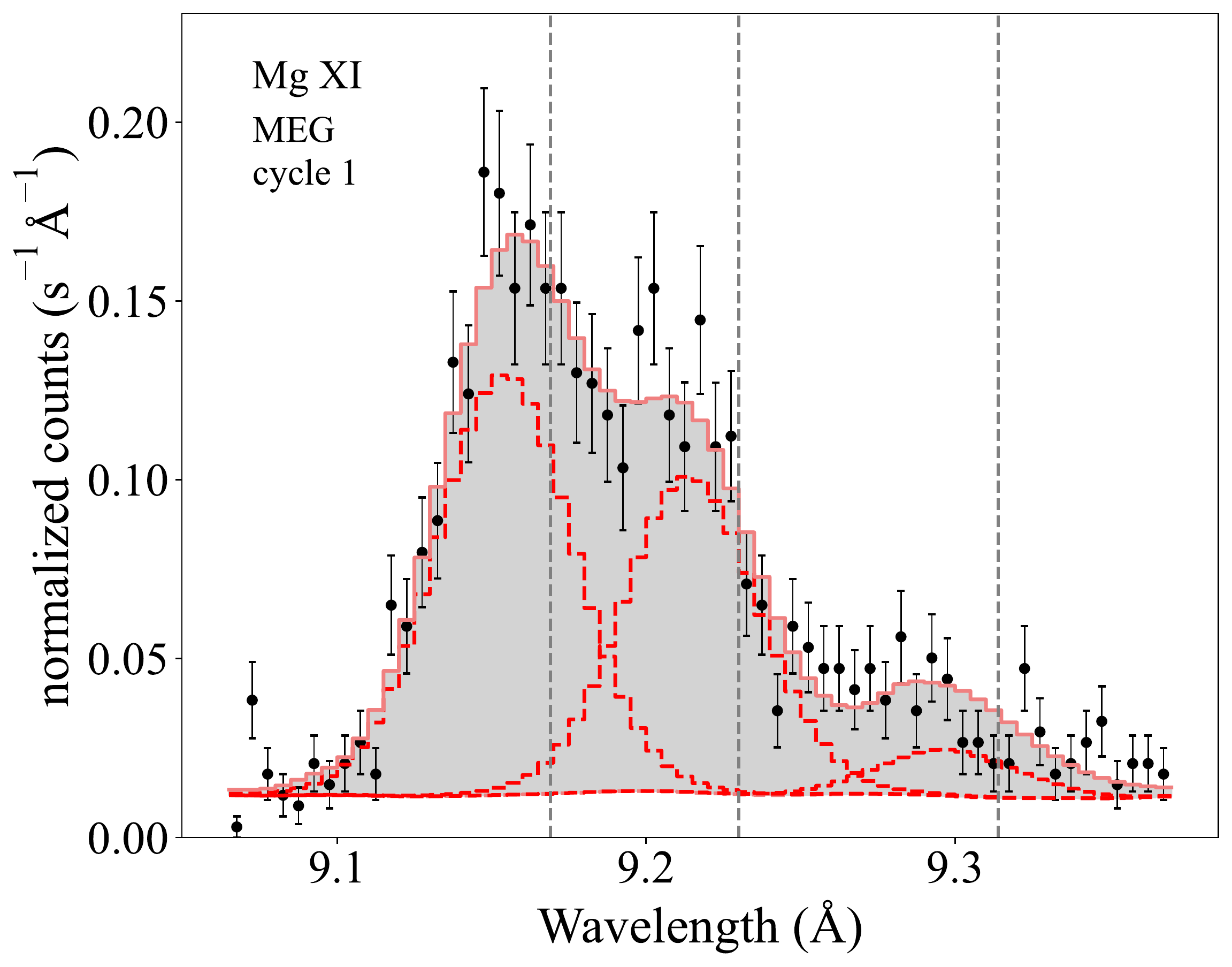}
    \includegraphics[angle=0,width=0.47\textwidth]{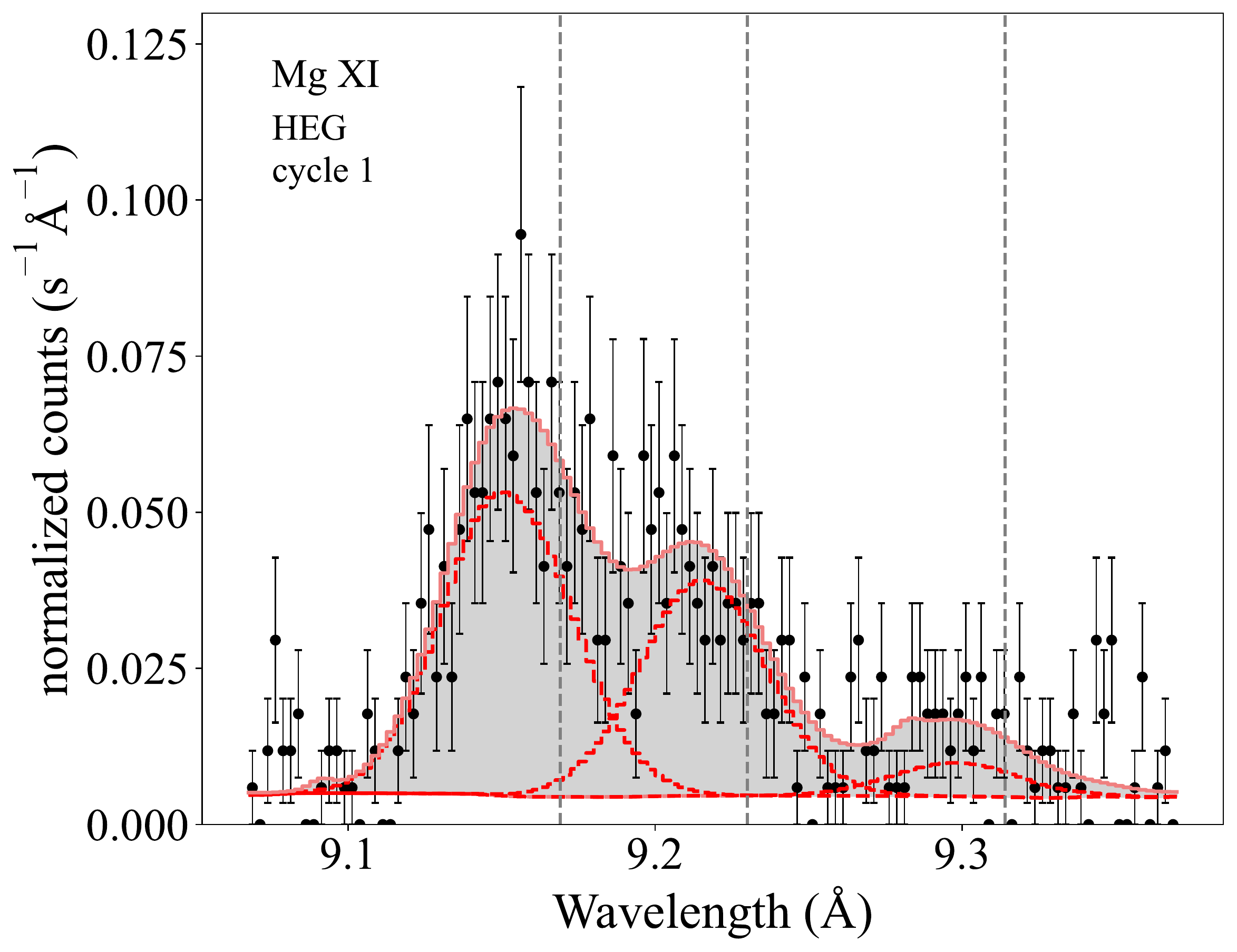}
    
    \caption{The \meg\/ (left) and \heg\/ (right) cycle 1 measurements of the S {\sc xv}, Si {\sc xiii}, and Mg {\sc xi} line complexes are shown as points with Poisson error bars, along with the best-fit {\it hegauss} models (solid red histogram). The three constituent Gaussian profiles for each best-fit model are shown via the dashed, darker red histograms and the continuum models fit to regions on either side of the line complexes are shown as the dashed red lines. The continuum-subtracted line complex fluxes are indicated by the grey shaded regions. The rest wavelengths of the three lines in each complex are indicated by vertical dashed lines. Modest blue shifts can be seen, especially in the Mg {\sc xi} complex, due to wind absorption. We note that the contributions from the modeled DR satellite lines can be seen, in the aggregate, via the gap between the individual three Gaussian components and the overall model flux. The shaded grey region thus includes both the He-like and the satellite line fluxes (but not the continuum). The contributions from individual satellite lines are shown and discussed in Appendix \ref{appendix:satellites}. 
    }
	\label{fig:hegauss1}
\end{figure*}

\begin{figure*}
\centering
    \includegraphics[angle=0,width=0.47\textwidth]{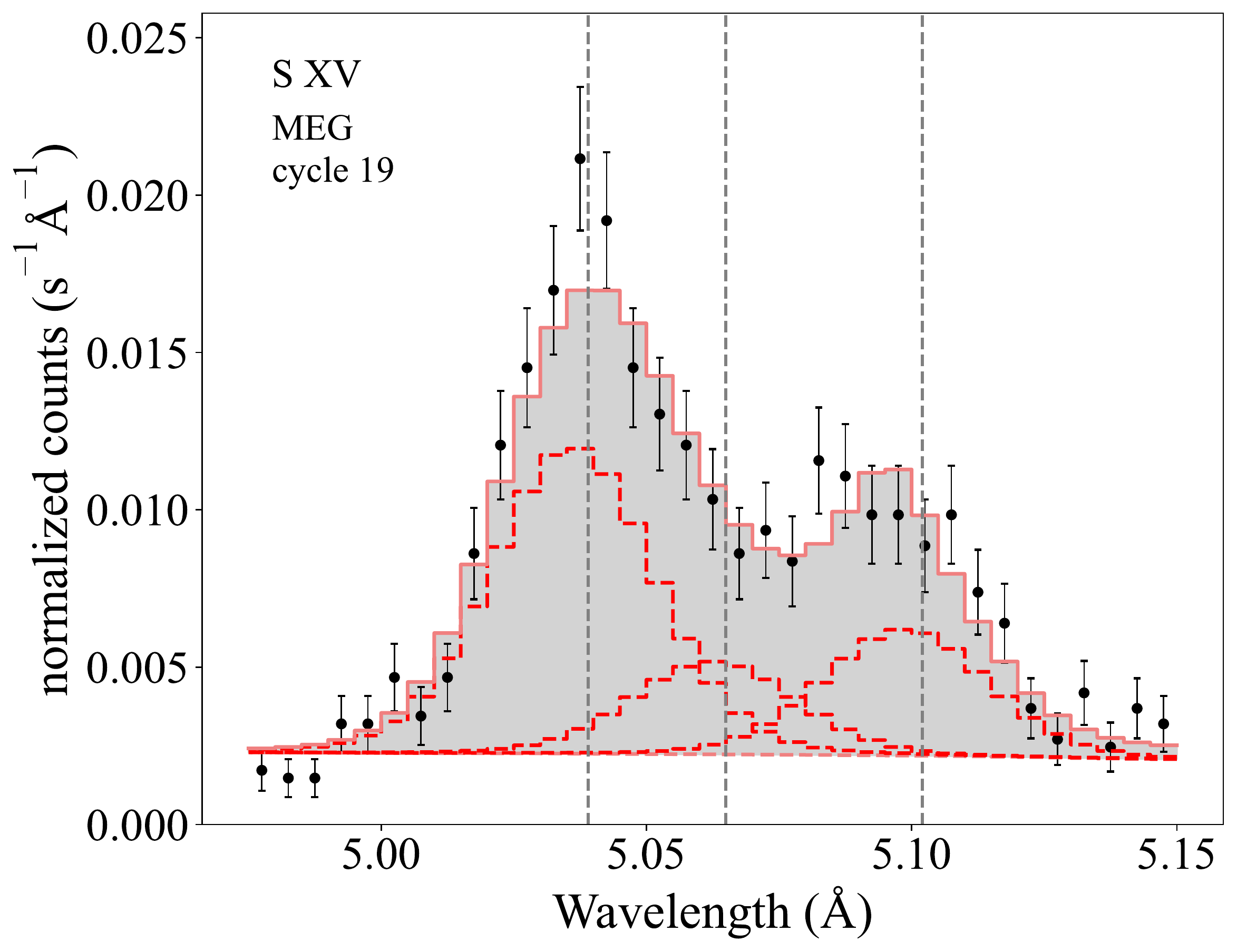}
    \includegraphics[angle=0,width=0.47\textwidth]{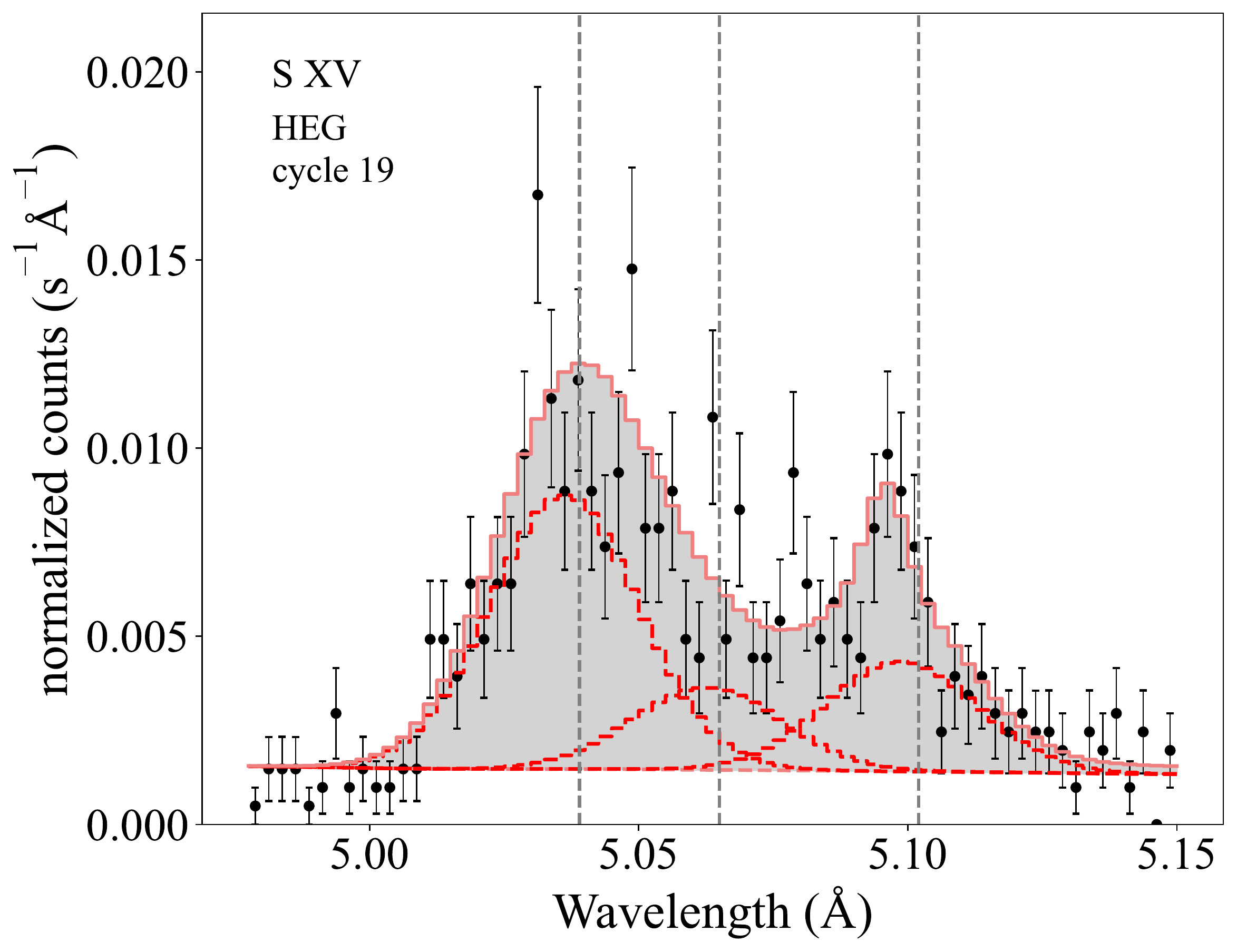}
    \includegraphics[angle=0,width=0.47\textwidth]{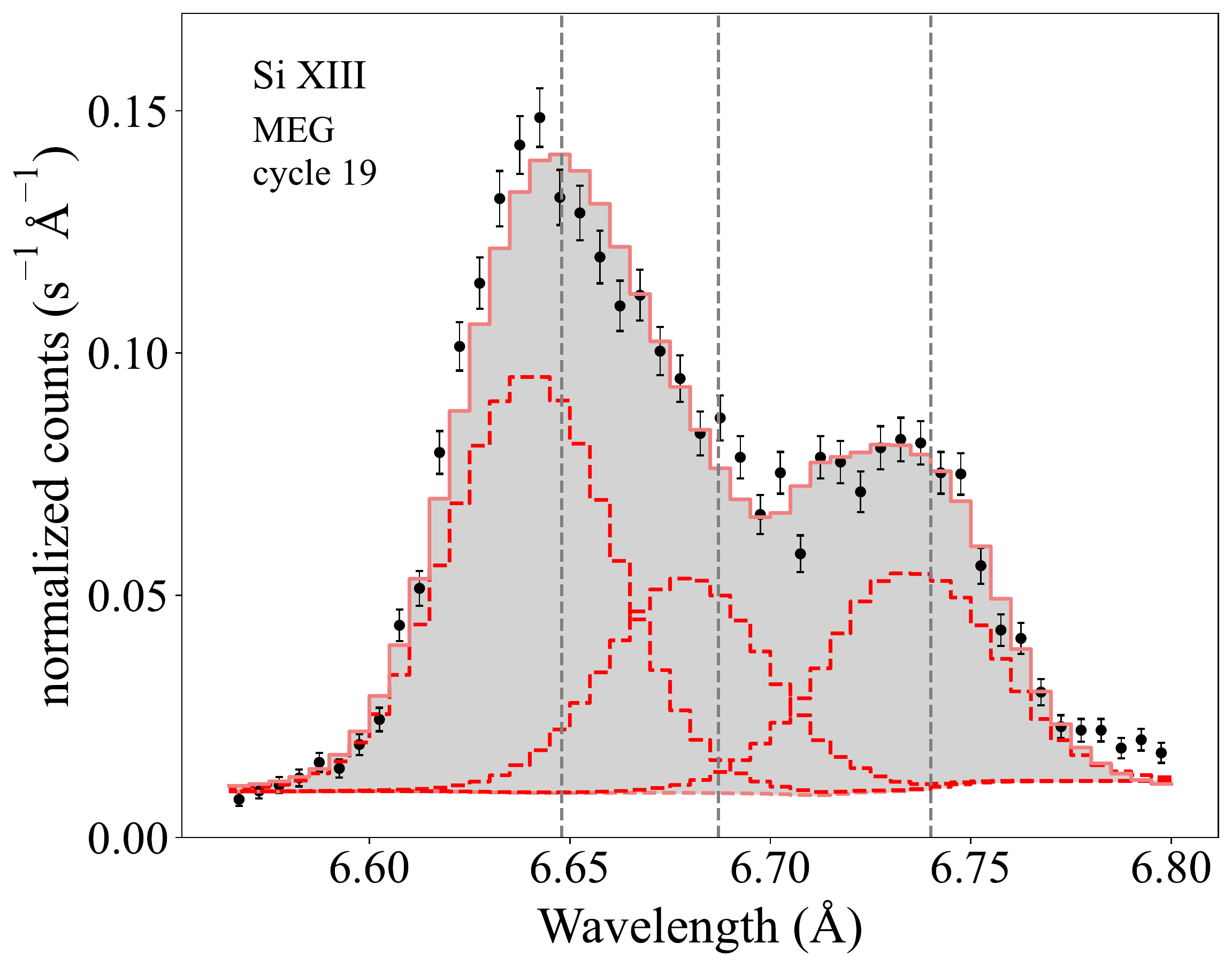}
    \includegraphics[angle=0,width=0.47\textwidth]{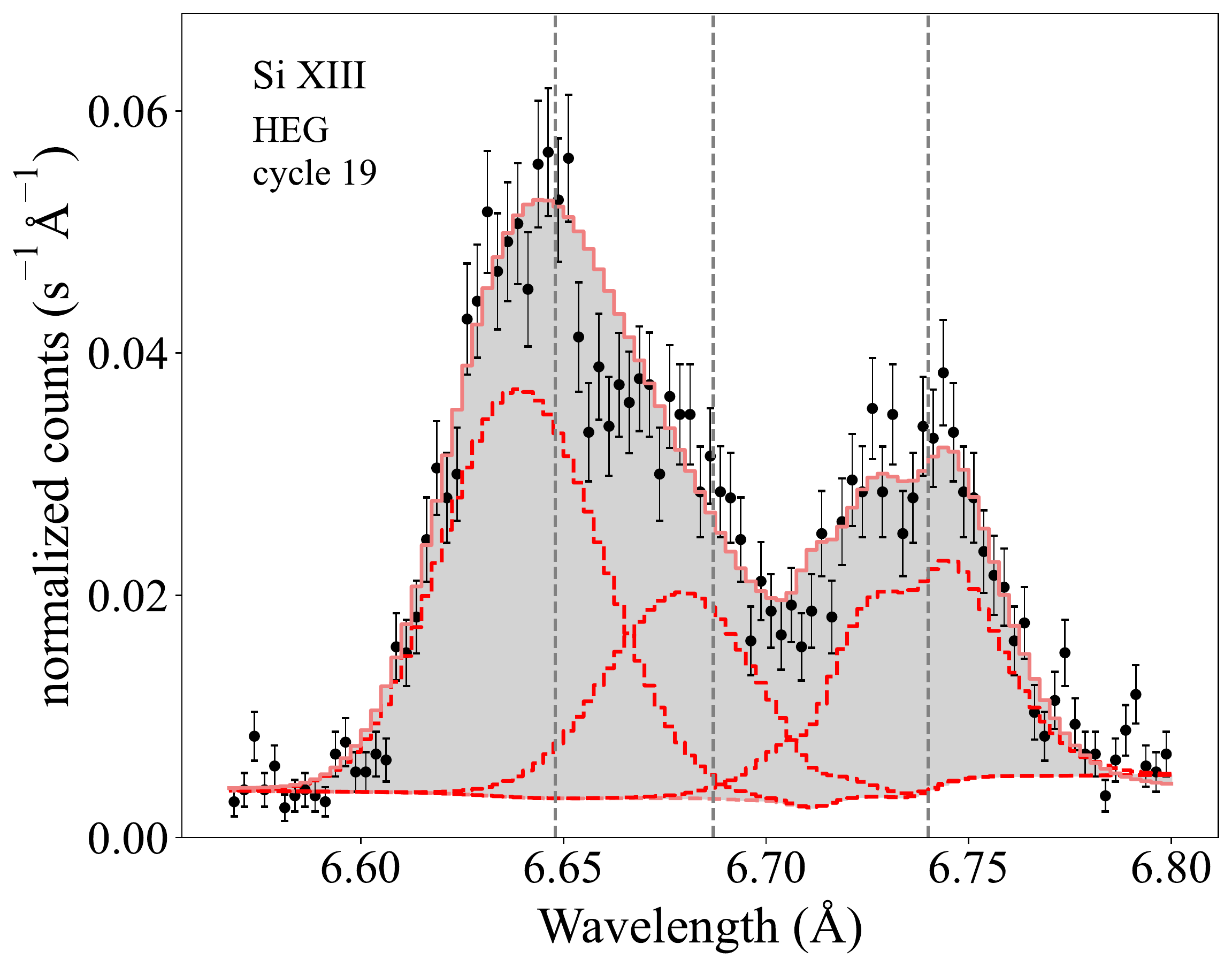}
    \includegraphics[angle=0,width=0.47\textwidth]{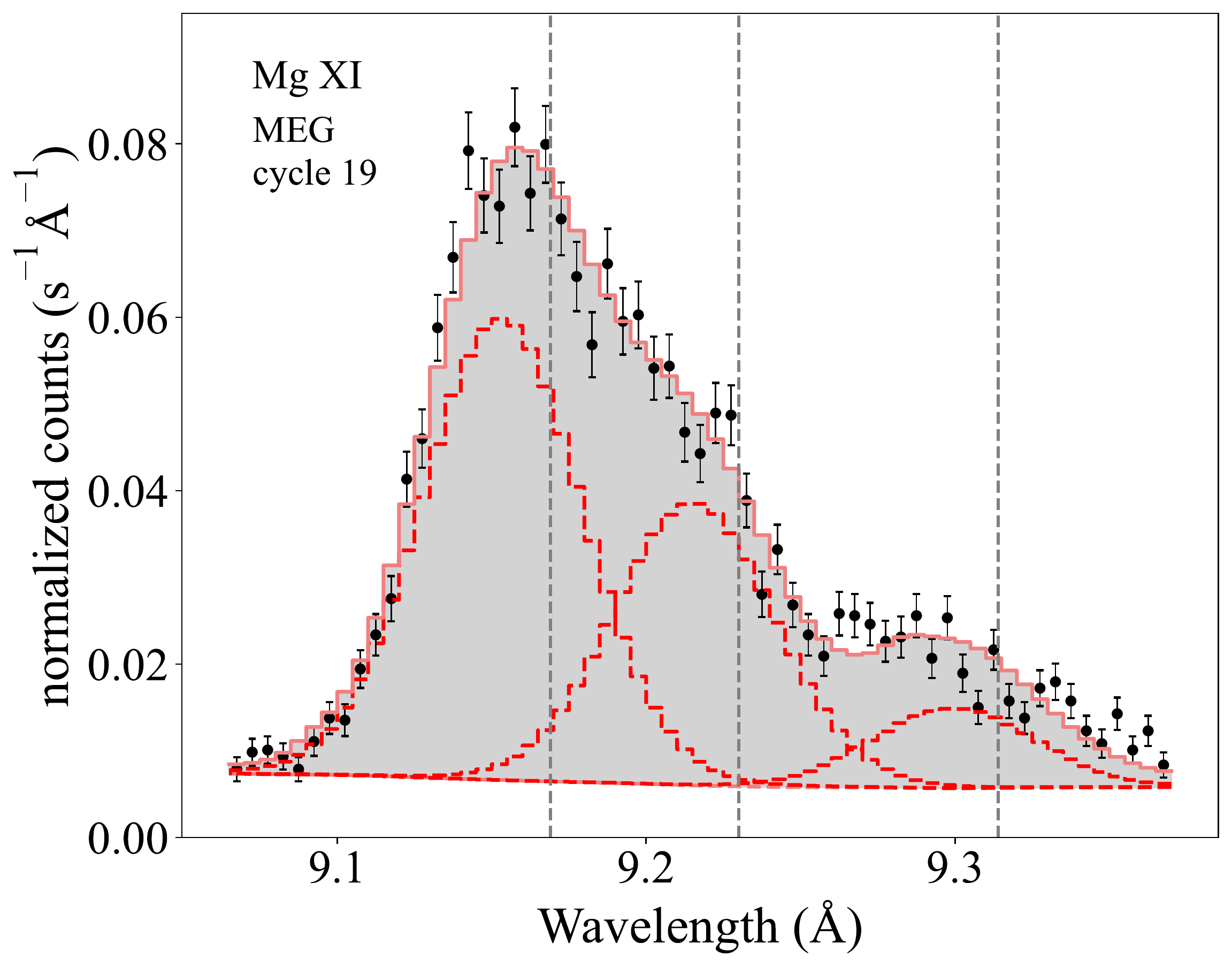}
    \includegraphics[angle=0,width=0.47\textwidth]{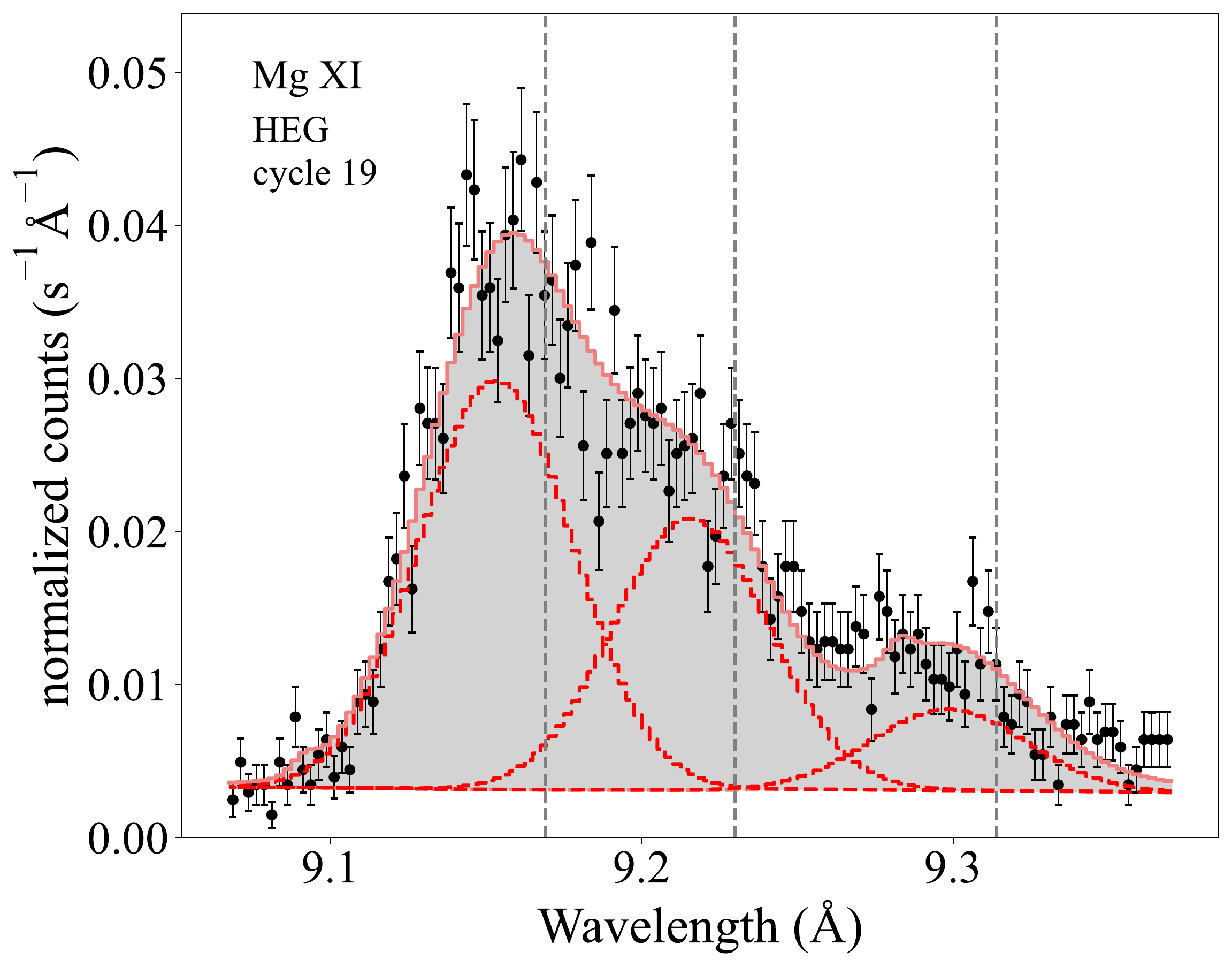}

    \caption{Same as Fig.\ \ref{fig:hegauss1}, but showing the {\it hegauss} fits to the cycle 19 data. Note that the 21 observations comprising these data were fit simultaneously but as separate data sets. They are co-added here for the purpose of clarity in display.}
	\label{fig:hegauss19}
\end{figure*}

\begin{figure}
\centering
 \vspace{-0.2in}
 \includegraphics[angle=0,width=0.38
 \textwidth]{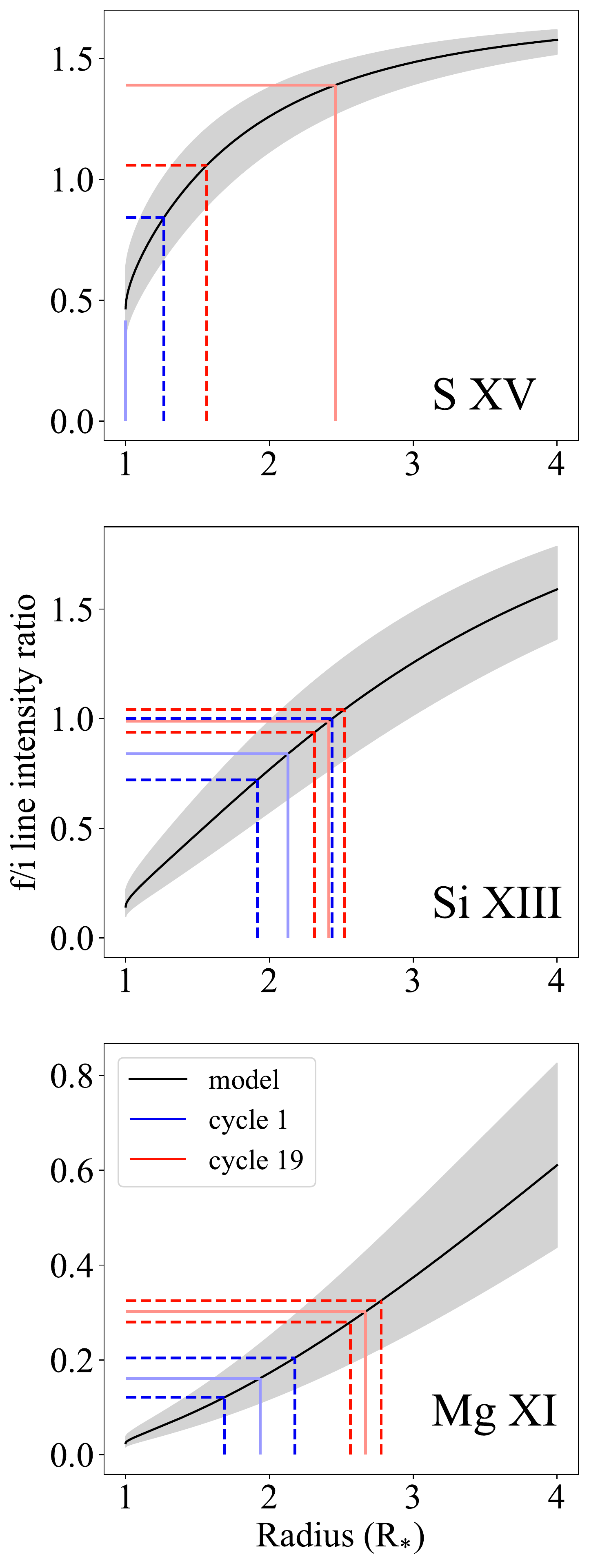}
	\caption{The formation radii, $R_{\rm f/i}$, inferred from the {\it hegauss} fitting are shown for each ion and both datasets (cycle 1 in blue and cycle 19 in red). The black curve in each plot represents the $f/i$ ratio's computed dependence on the location of the X-ray emitting plasma (equation \ref{eq:Rofr}) and the grey band surrounding it represents the results for the assumed uncertainty in the photospheric flux of $1.5H_{\nu}$ and $0.67H_{\nu}$. The {\it hegauss} measurements are indicated by the colored solid horizontal lines, and the perpendiculars dropped to the x-axis show the inferred formation radii. The dotted blue and red lines indicate the 68 per cent statistical error on the line ratios and, along the x-axis, on the formation radii. For S {\sc xv} the upper confidence limit for the cycle 19 data is not shown because it exceeds the no-photoexcitation limit, $\mathcal{R}_{\rm o}$, and so it and the corresponding maximum radial location is unconstrained by the data, while the lower limit for the cycle 1 data is not shown because it is consistent with the surface of the star. Thus the red dashed line in the top panel represents the lower confidence limit on \Rfir\/ in cycle 19 and the dashed blue line represents the upper confidence limit in cycle 1. 
	}   
	\label{fig:projRfir}
\end{figure}

The results of the {\it hegauss} modeling for both cycle 1 and cycle 19 are presented in Table \ref{tab:hegauss} and the best fits to the data are shown in Fig.\ \ref{fig:hegauss1} and Fig.\ \ref{fig:hegauss19}, respectively. We model the radial dependence of the $f/i$ ratios measured from the {\it hegauss} fits and list the results in Tables \ref{tab:Rfir_values1} and \ref{tab:Rfir_values19}, with the results shown graphically in Fig.\ \ref{fig:projRfir}. To conservatively include systematic model atmosphere uncertainties in the UV fluxes, we show results for higher and lower UV flux cases (scaled by 1.5 and 0.67, respectively) in addition to the fluxes adopted from the favored, $T_{\rm eff} = 40,000$ K, $\log g = 3.50$, \tlusty\/ model. These are included in the two tables and also included in Fig.\ \ref{fig:projRfir}, as a grey band around the solid line showing $\mathcal{R}(r)$ derived from the adopted atmosphere model. 

The fits to the data in all six cases (three line complexes in two different sets of observations) are formally good. With the much larger combined exposure time of the cycle 19 observations, the uncertainties on the model parameters are smaller for those more recently taken data. The $f/i$ ratios, line fluxes, and velocity widths have increased modestly, but significantly, between cycle 1 and cycle 19. 

\begin{figure*}
\centering
    \includegraphics[angle=0,width=0.47\textwidth]{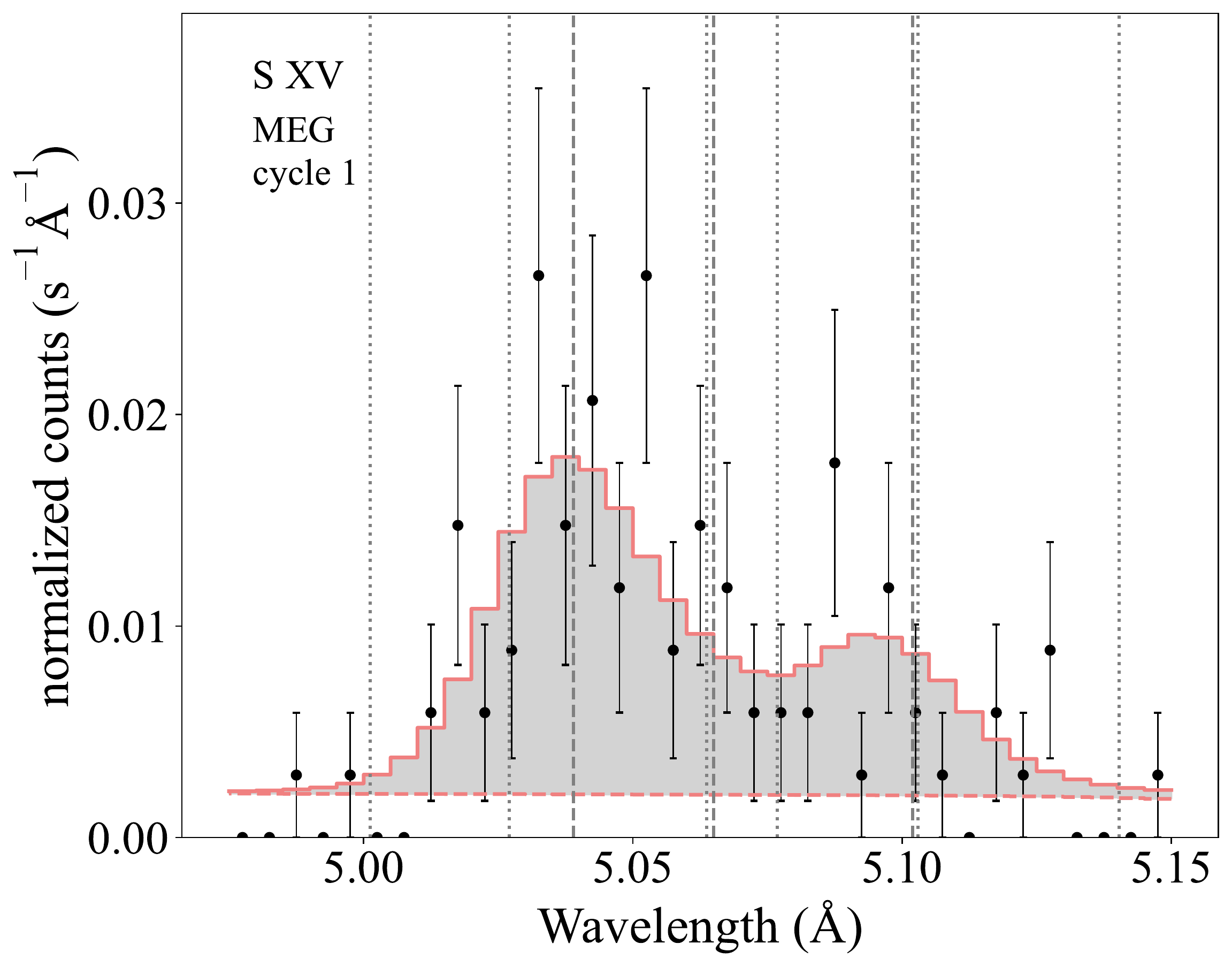}
    \includegraphics[angle=0,width=0.47\textwidth]{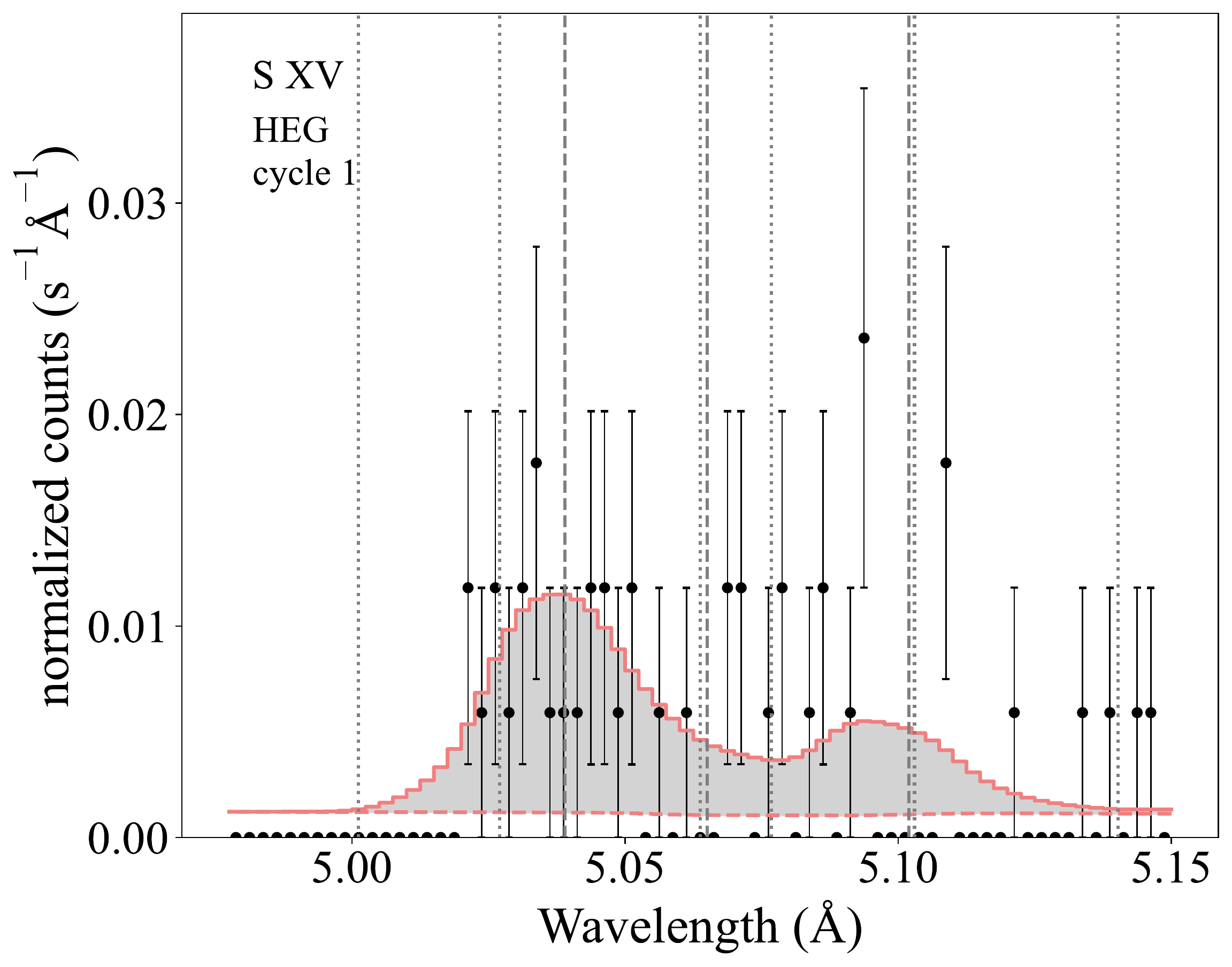}
    \includegraphics[angle=0,width=0.47\textwidth]{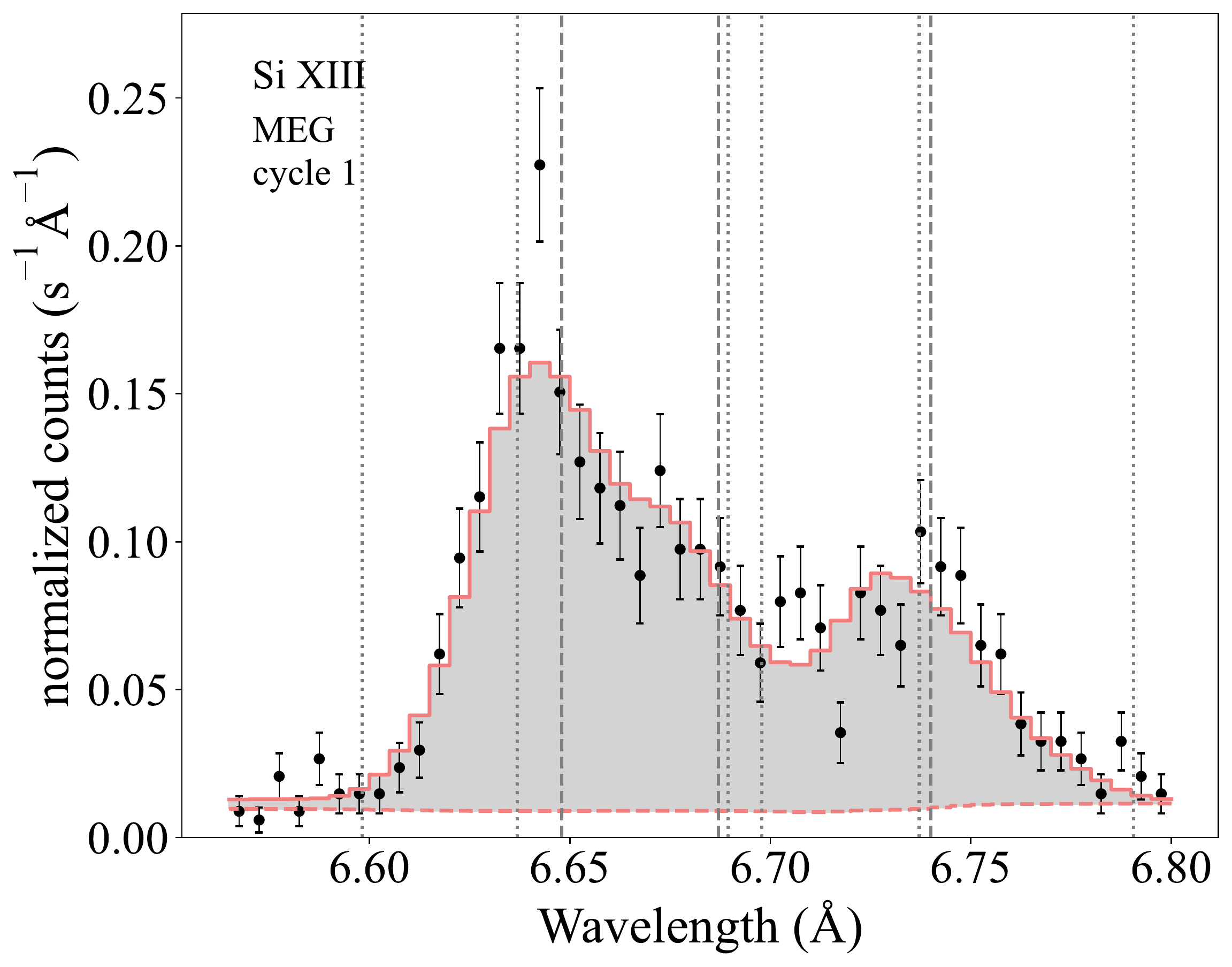}
    \includegraphics[angle=0,width=0.47\textwidth]{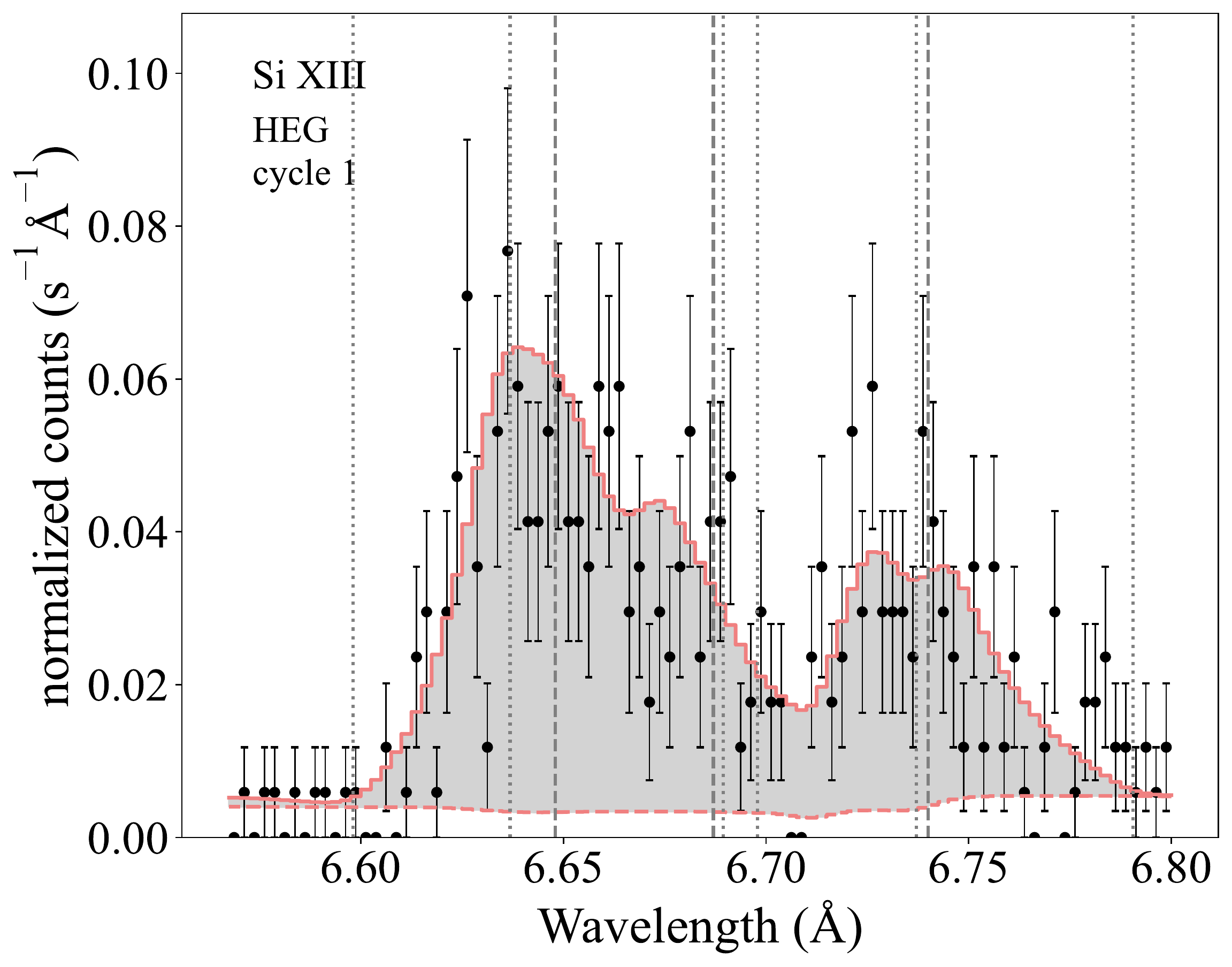}
    \includegraphics[angle=0,width=0.47\textwidth]{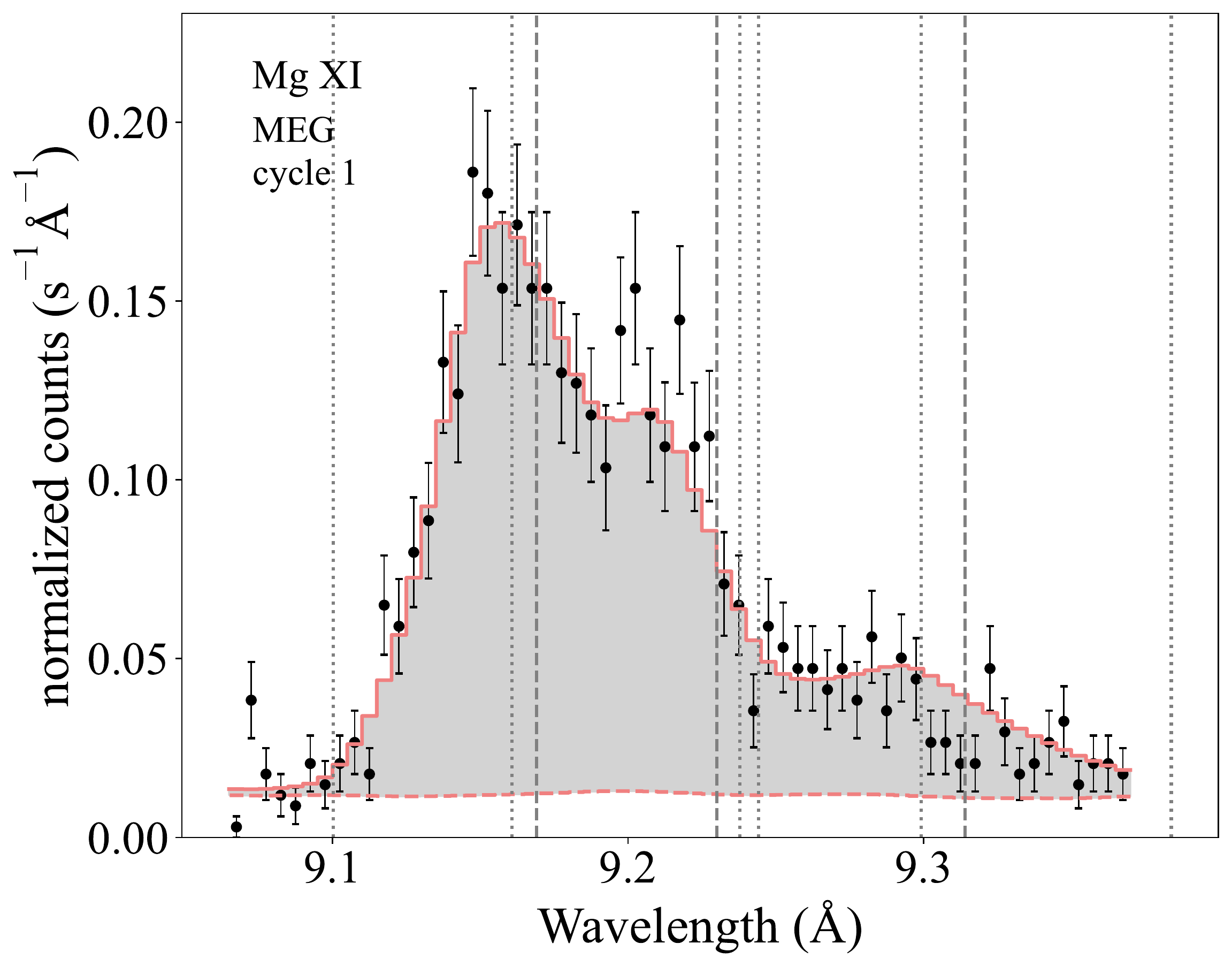}
    \includegraphics[angle=0,width=0.47\textwidth]{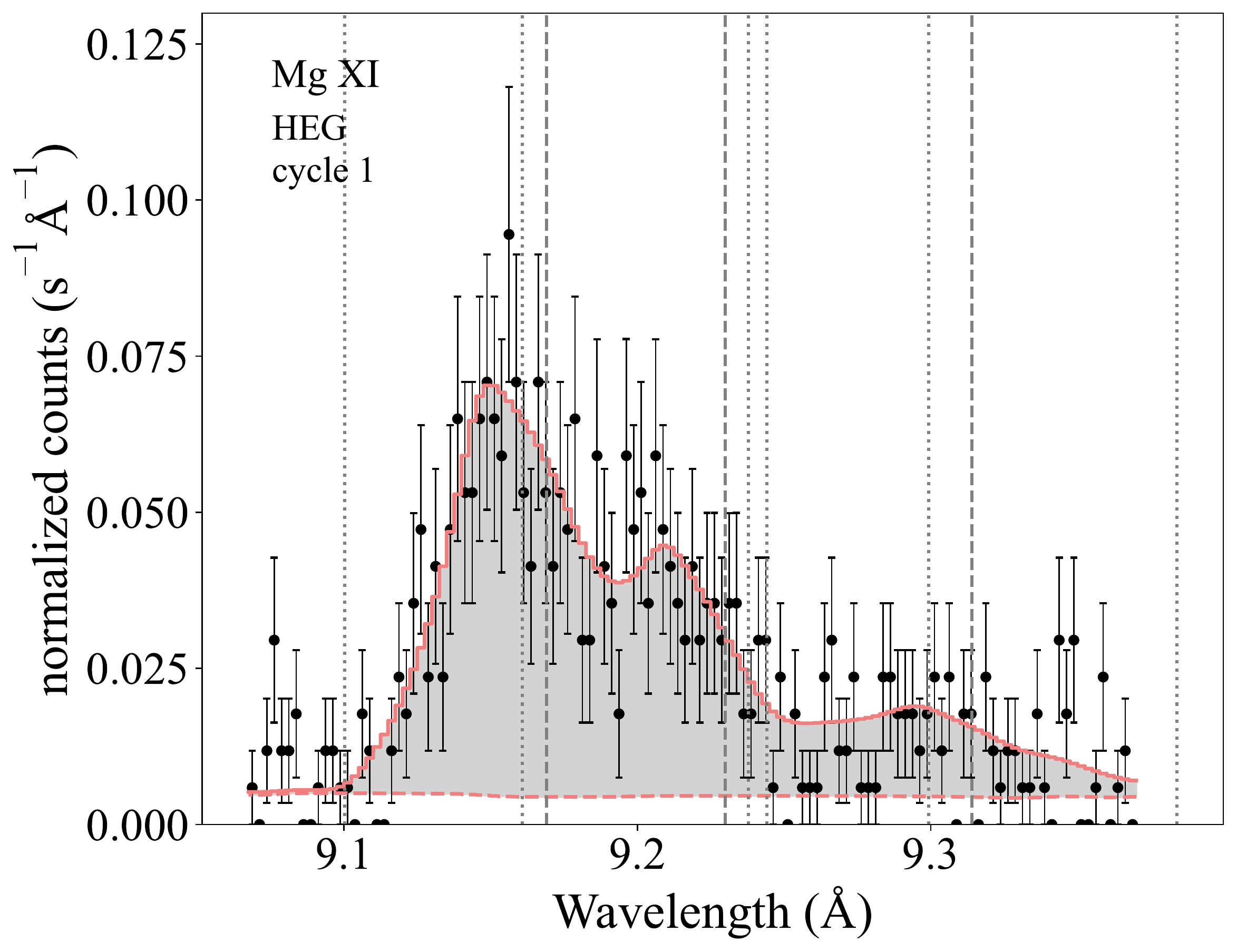}
    
    \caption{The \meg\/ (left) and \heg\/ (right) cycle 1 measurements of the S {\sc xv}, Si {\sc xiii}, and Mg {\sc xi} line complexes are shown as points with Poisson error bars, along with the best-fit {\it hewind} models, including also the satellite contributions (red histogram). The continuum is indicated by the dashed red line and the continuum-subtracted line complex (including modeled DR satellite lines) fluxes are indicated by the grey shaded regions. The rest wavelengths of the three lines in each complex are indicated by vertical dashed lines while the minimum and maximum wind Doppler shifted wavelengths, determined by the wind terminal velocity of 2250 \kms, are indicated by the vertical dotted lines -- one on each side of each dashed line.}
	\label{fig:hewind1}
\end{figure*}

\begin{figure*}
\centering
    \includegraphics[angle=0,width=0.47\textwidth]{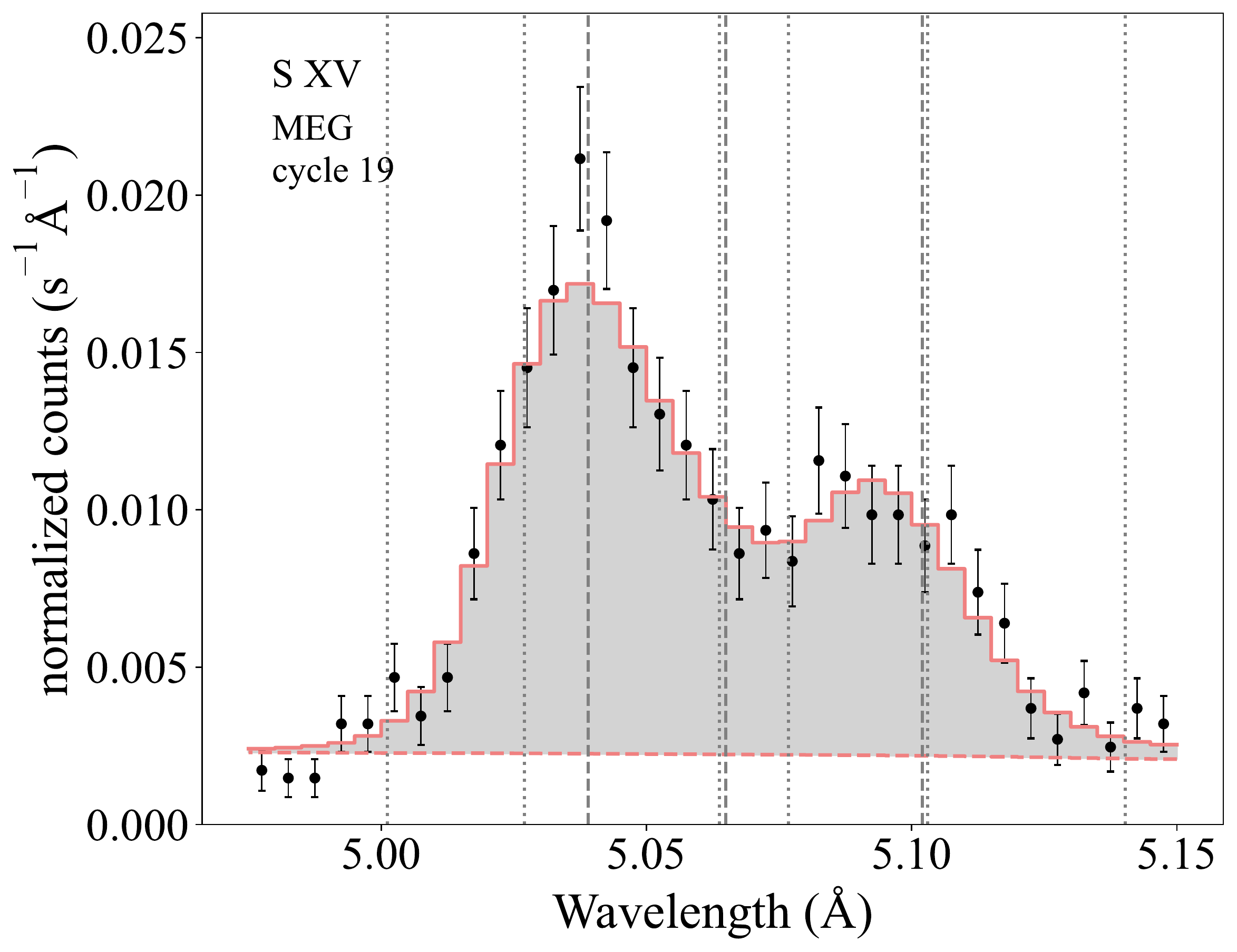}
    \includegraphics[angle=0,width=0.47\textwidth]{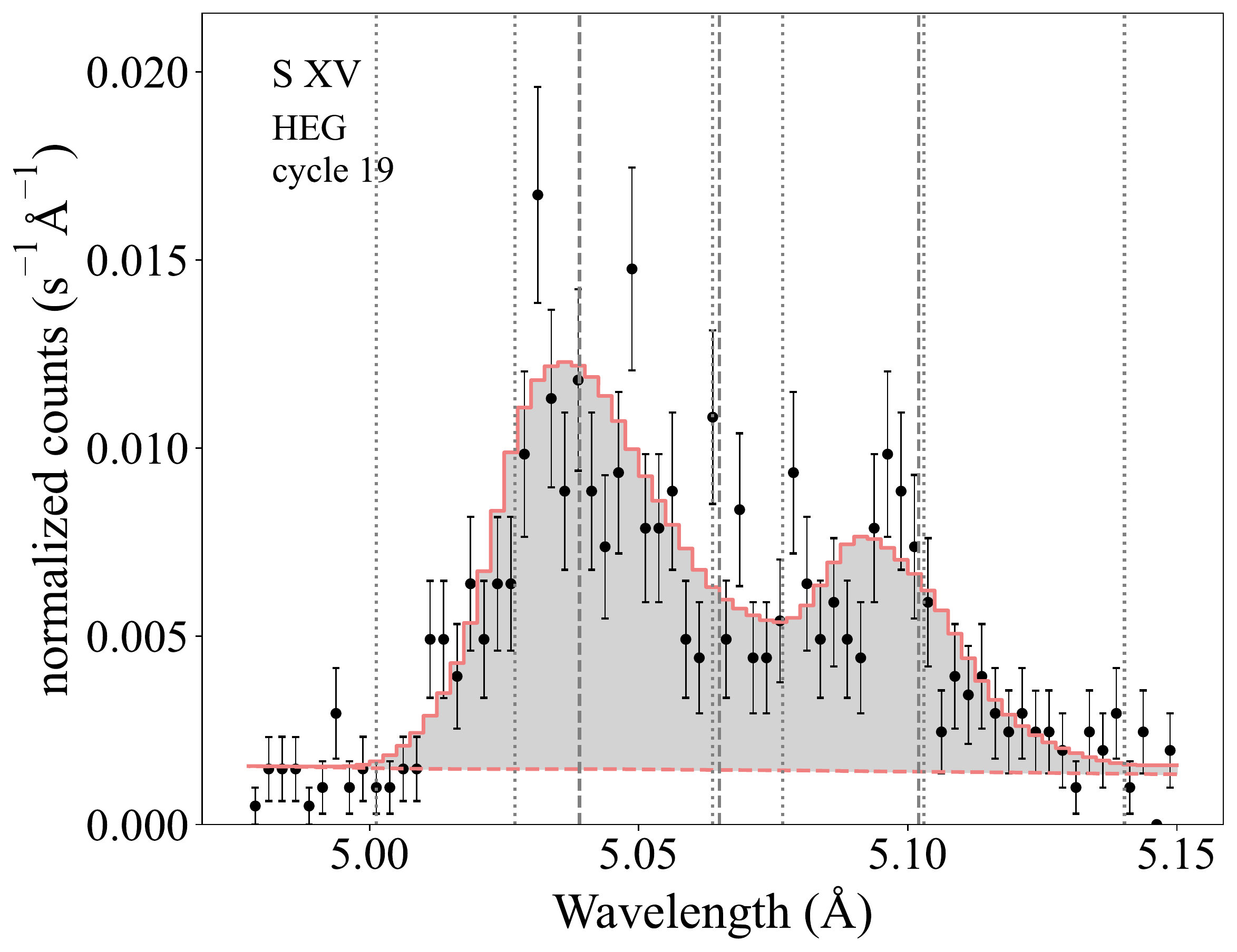}
    \includegraphics[angle=0,width=0.47\textwidth]{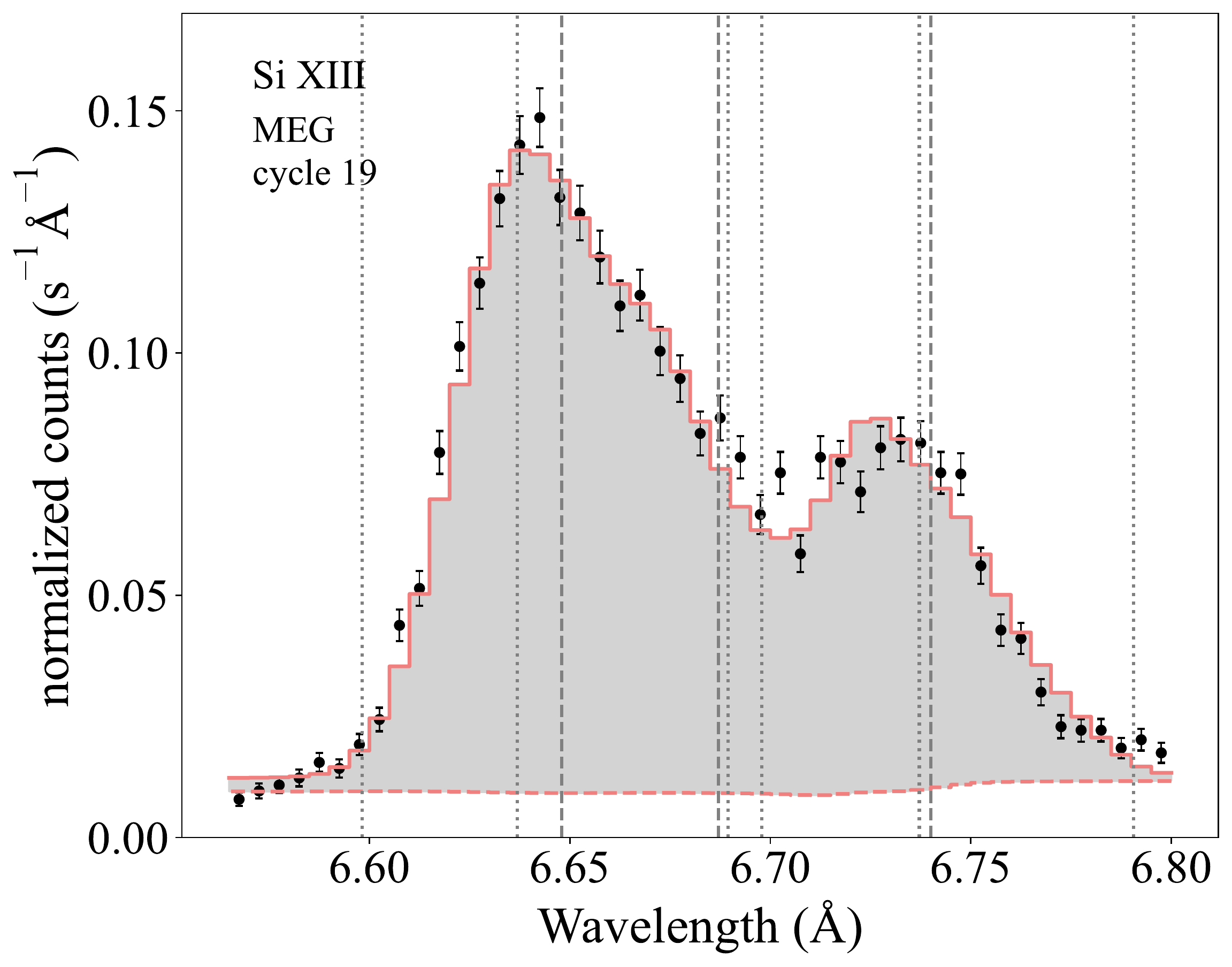}
    \includegraphics[angle=0,width=0.47\textwidth]{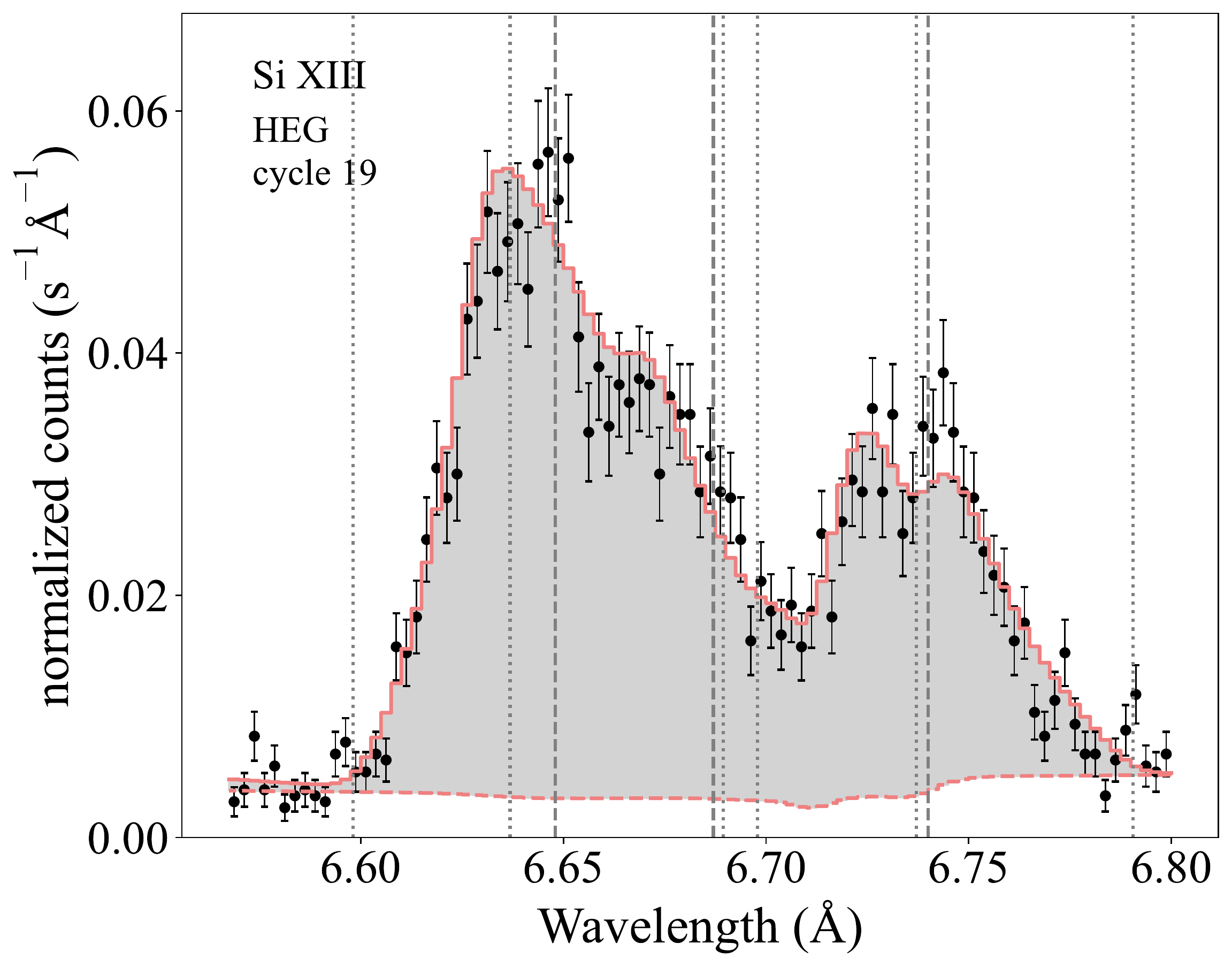}
    \includegraphics[angle=0,width=0.47\textwidth]{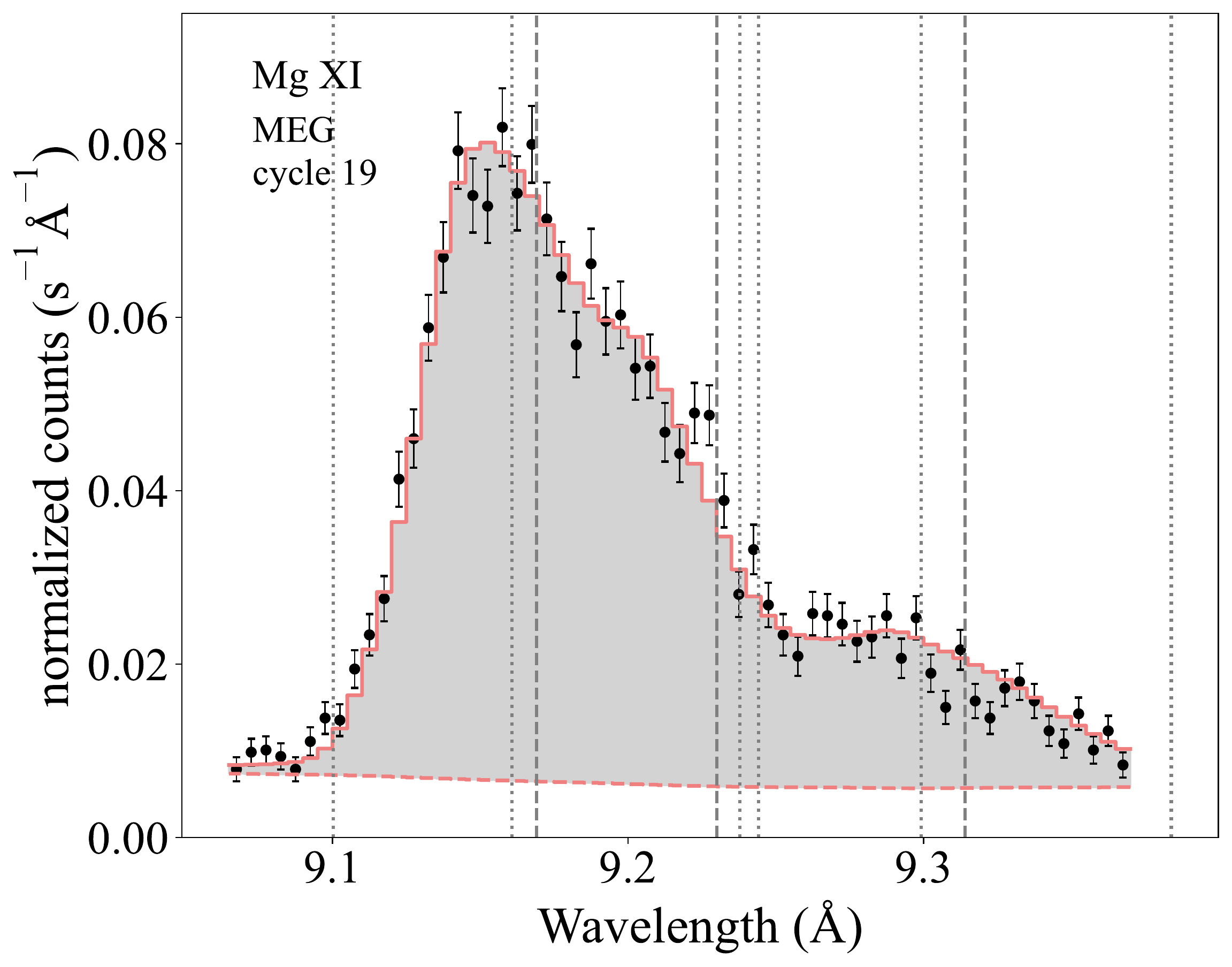}
    \includegraphics[angle=0,width=0.47\textwidth]{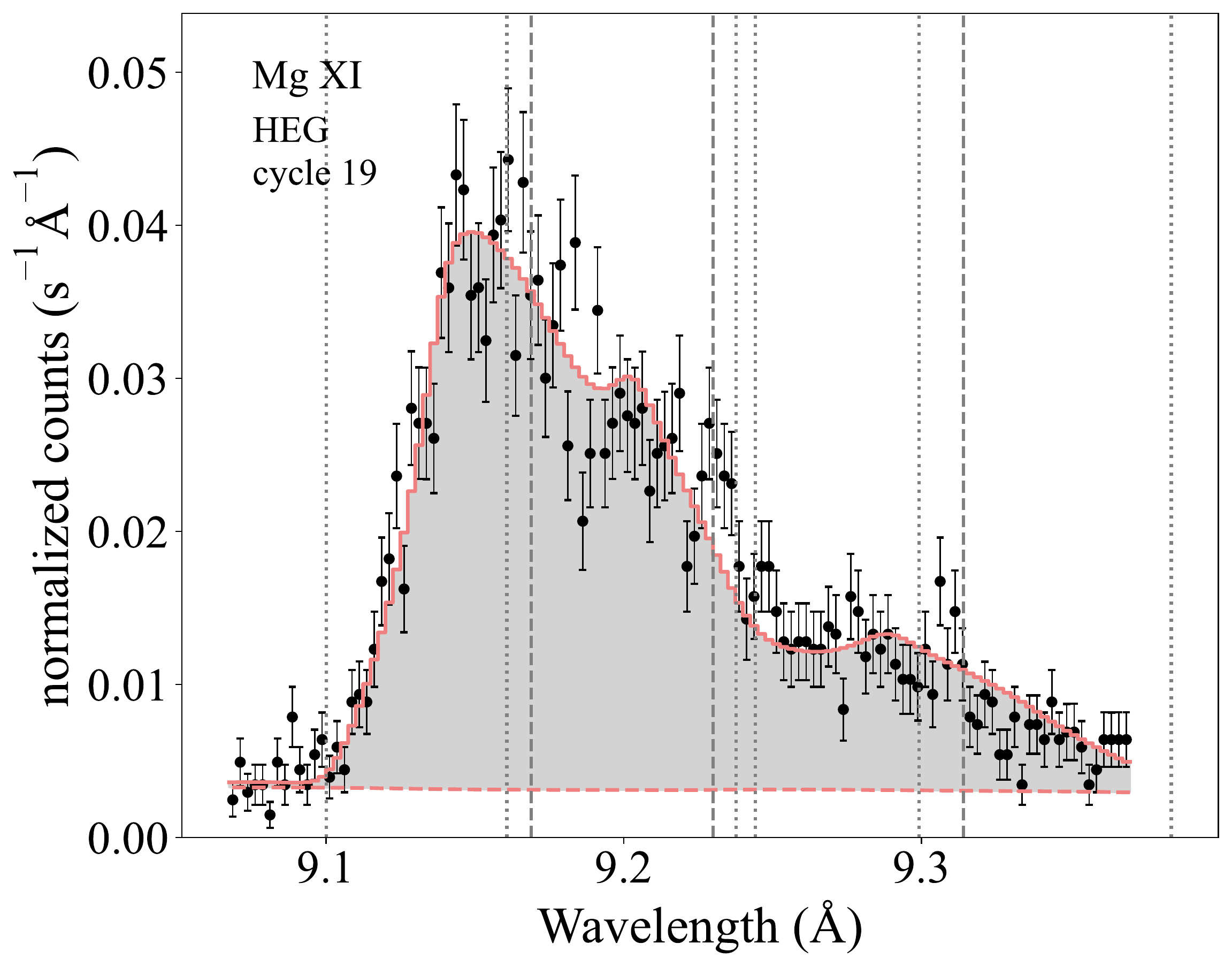}
    \caption{Same as Fig.\ \ref{fig:hewind1}, but showing the {\it hewind} fits to the cycle 19 data.}
	\label{fig:hewind19}
\end{figure*}

\begin{figure}
\centering
 \includegraphics[angle=0,width=0.45\textwidth]{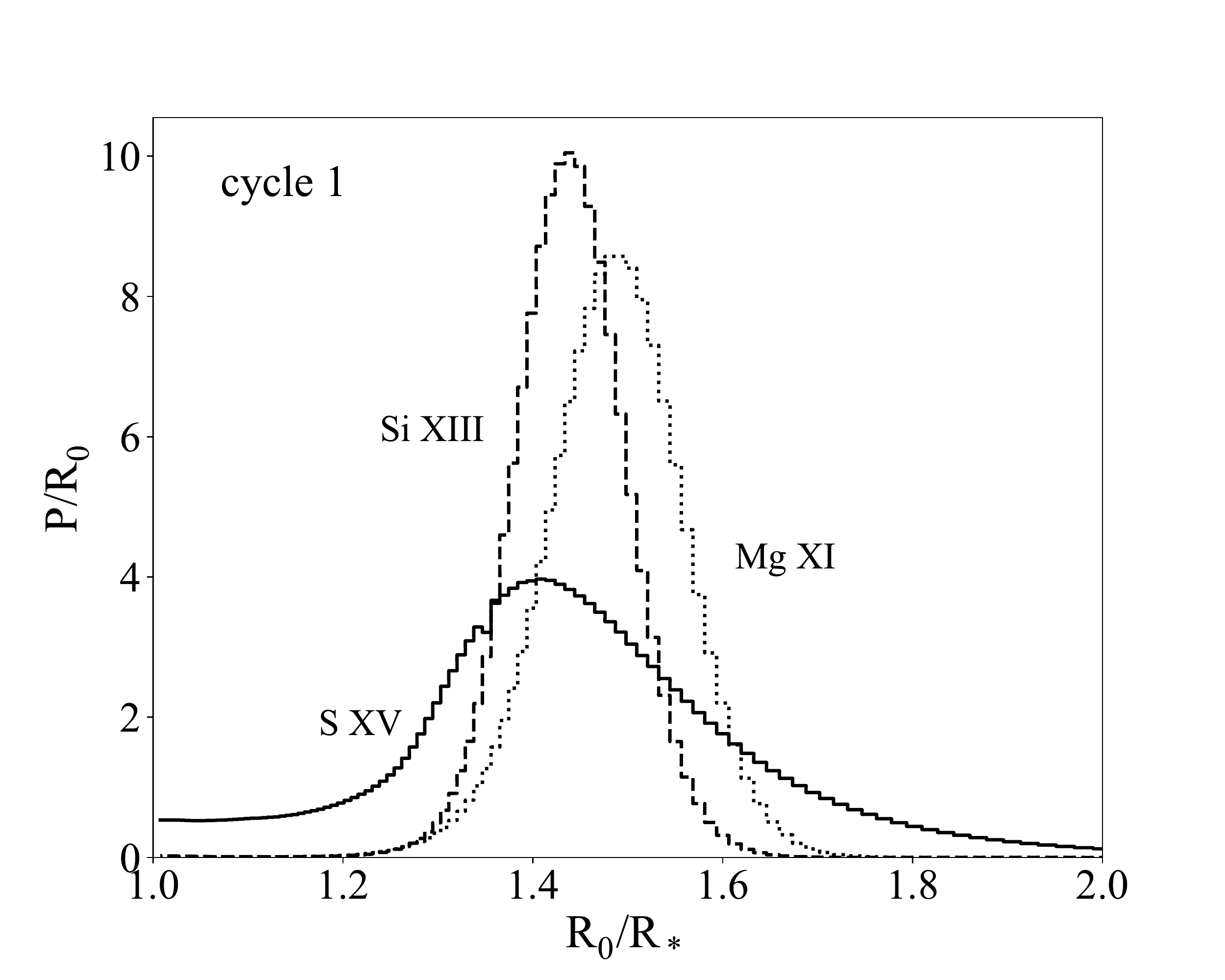}
 \includegraphics[angle=0,width=0.45\textwidth]{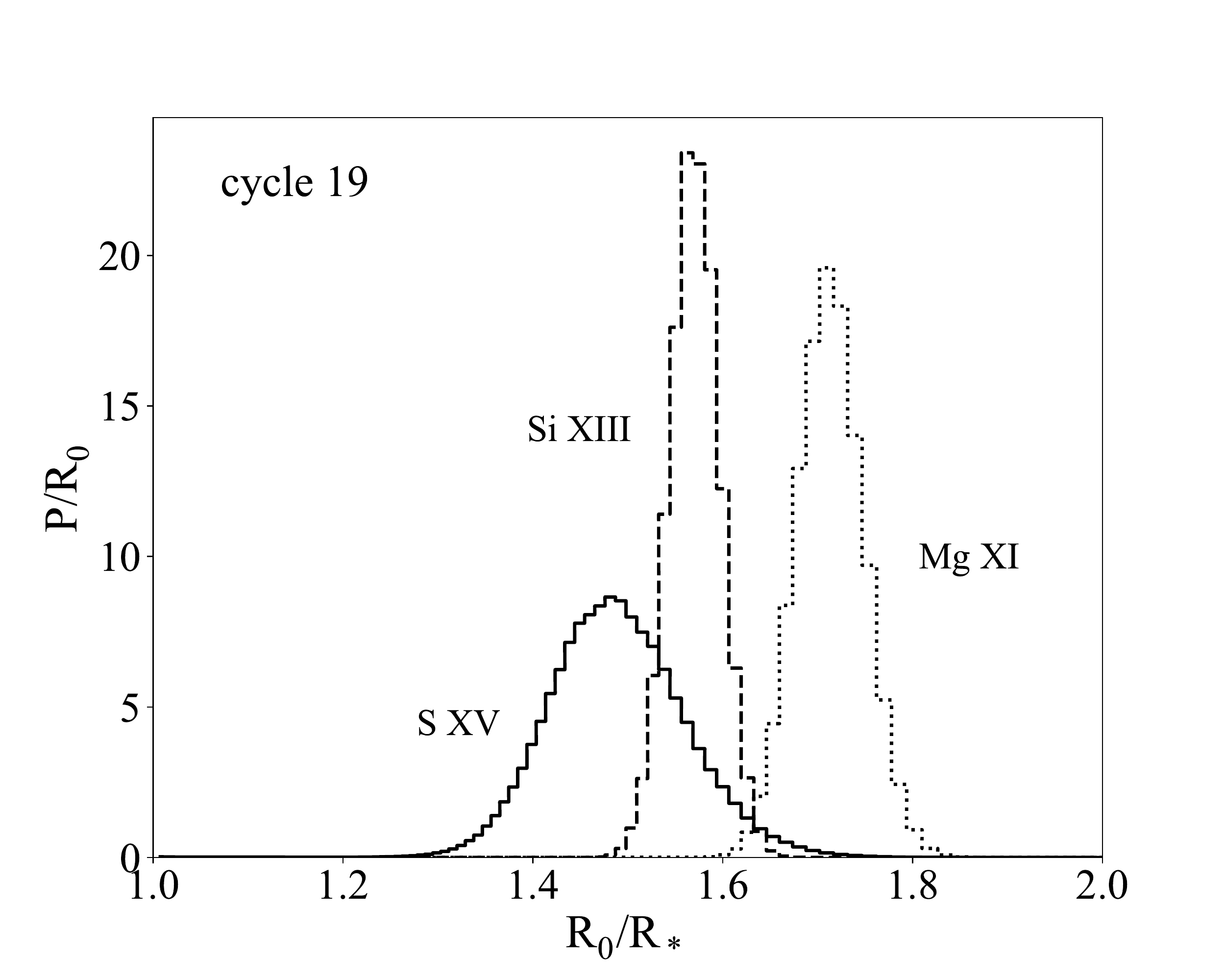}
	\caption{The probability distributions (per stellar radius) -- S {\sc xv} (solid), Si {\sc xiii} (dashed), and Mg {\sc xi} (dotted) -- derived for the onset radius parameter, \Ro, of the {\it hewind} modeling (cycle 1 top and cycle 19 bottom). We emphasize that these probabilities are for the onset radius and that the emission in this model is assumed to extend throughout the wind above \Ro\/ with a weighting proportional to the square of the local wind density. 
	}   
	\label{fig:Ro_probs}
\end{figure}

The results of the {\it hewind} modeling for both cycle 1 and cycle 19 are presented in Table \ref{tab:hewind}. The fits are shown graphically in Fig.\ \ref{fig:hewind1}, for cycle 1 data, and Fig.\ \ref{fig:hewind19} for cycle 19 data. In most cases, this model provides statistically better fits to the already adequate Gaussian line-shape fits of {\it hegauss}. Note that whereas that model fits the $f/i$ ratio directly, the free parameters of {\it hewind} are a characteristic optical depth, \taustar, and a shock onset radius, \Ro. The probability distributions of the key parameter -- \Ro\/ -- are shown in Fig.\ \ref{fig:Ro_probs}. These are based on the run of $\Delta{C}$ with \Ro, according to $P \propto e^{-{\Delta}C/2}$, and in the figure (as well as in Table \ref{tab:hewind}) the tighter constraints provided by the cycle 19 data, as well as the modest, but significant, changes between the two cycles, are evident. We stress here that in the {\it hewind} model \Ro\/ is the onset radius of X-ray emission with emission also generated throughout the wind above \Ro. We also reiterate that all the modeling presented here accounts for the contamination of the He-like complexes by satellite line emission via our inclusion of dozens of satellite lines in the spectral fitting, as described in Appendix \ref{appendix:satellites}.

\begin{table*}
  \caption{Best-fit {\it hegauss} parameters for S {\sc xv}, Si {\sc xiii}, and Mg {\sc xi}}
  \vspace{1mm}
\centering
\def\arraystretch{1.4}
\begin{tabular}{cccc}
  \hline 
  Cycle 1 & S {\sc xv} & Si {\sc xiii} & Mg {\sc xi} \\
  \hline 
  $\mathcal{R}$ & $0.42_{-0.27}^{+0.42}$ & $0.84_{-0.12}^{+0.15}$ & $0.16_{-0.04}^{+0.04}$ \\
  $\mathcal{G}$ & $1.10_{-0.50}^{+0.79}$ & $0.98_{-0.08}^{+0.12}$ & $0.82_{-0.85}^{+0.92}$ \\
  $\sigma_v$ (km s$^{-1}$) & $546_{-123}^{+139}$ & $682_{-39}^{+43}$ & $663_{-32}^{+36}$ \\
  $\delta_v$ (km s$^{-1}$) & $-315_{-192}^{+140}$ & $-272_{-46}^{+45}$ & $-510_{-49}^{+48}$ \\
  {\it hegauss} norm (photons cm$^{-2}$ s$^{-1}$) & $1.82_{-0.20}^{+0.20}\times 10^{-5}$ & $9.11_{-0.27}^{+0.28}\times 10^{-5}$ & $1.50_{-0.05}^{+0.04}\times 10^{-4}$ \\
  {\it pow} norm (photons keV$^{-1}$ cm$^{-2}$ s$^{-1}$ at 1 keV) & $7.33_{-0.66}^{+0.70}\times 10^{-4}$ & $1.12_{-0.06}^{+0.06}\times 10^{-3}$ & $1.84_{-0.08}^{+0.08}\times 10^{-3}$ \\
  \hline
  Cycle 19 & S {\sc xv} & Si {\sc xiii} & Mg {\sc xi} \\
  \hline 
  $\mathcal{R}$ & $1.39_{-0.33}^{+0.54}$ & $0.99_{-0.05}^{+0.05}$ & $0.30_{-0.02}^{+0.03}$ \\
  $\mathcal{G}$ & $0.74_{-0.12}^{+0.13}$ & $1.02_{-0.04}^{+0.04}$ & $0.88_{-0.04}^{+0.05}$ \\
  $\sigma_v$ (km s$^{-1}$) & $786_{-54}^{+57}$ & $837_{-14}^{+14}$ & $790_{-16}^{+17}$ \\
  $\delta_v$ (km s$^{-1}$) & $-147_{-55}^{+50}$ & $-338_{-14}^{+17}$ & $-495_{-24}^{+24}$ \\
  {\it hegauss} norm (photons cm$^{-2}$ s$^{-1}$) & $2.15_{-0.06}^{+0.07}\times 10^{-5}$ & $1.10_{-0.01}^{+0.01}\times 10^{-4}$ & $1.73_{-0.02}^{+0.02}\times 10^{-4}$ \\
  {\it pow} norm (photons keV$^{-1}$ cm$^{-2}$ s$^{-1}$ at 1 keV) & $8.97_{-0.22}^{+0.23}\times 10^{-4}$ & $1.38_{-0.02}^{+0.02}\times 10^{-3}$ & $2.17_{-0.03}^{+0.03}\times 10^{-3}$ \\
  \hline
\label{tab:hegauss}
\end{tabular}
\end{table*}

\clearpage 

\clearpage 

\begin{table}
\caption{
Formation radii based on the {\it hegauss} fitting of the cycle 1 data 
}
\centering
\def\arraystretch{1.4}
\begin{tabular}{cccc}
  \hline
  S {\sc xv} & min & best & max \\
  \hline
  $0.67*H_\nu$ & 1.00 & 1.00 & 1.11 \\
  $H_\nu$ & 1.00 & {\bf 1.00} & 1.27 \\
  $1.5*H_\nu$ & 1.00 & 1.03 & 1.49 \\ 
  \hline
  Si {\sc xiii} &  &  &  \\
  \hline
  $0.67*H_\nu$ & 1.60 & 1.77 & 2.02 \\
  $H_\nu$ & 1.92 & {\bf 2.13} & 2.44 \\
  $1.5*H_\nu$ & 2.31 & 2.58 & 2.96 \\
  \hline
  Mg {\sc xi} &  &  &  \\
  \hline
  $0.67*H_\nu$ & 1.42 & 1.61 & 1.81 \\
  $H_\nu$ & 1.69 & {\bf 1.93} & 2.18 \\
  $1.5*H_\nu$ & 2.03 & 2.34 & 2.64 \\
  \hline
\end{tabular}
\\
{
The single formation radii, \Rfir, in units of the stellar radius are derived from equation \ref{eq:Rofr} and the values of $\mathcal{R} = f/i$ from the {\it hegauss} fitting (shown graphically in Fig.\ \ref{fig:projRfir}). For each of the three line complexes, ``min'' and ``max'' indicate the 68\% confidence limits of the derived formation radii. The results for the assumed higher and lower UV fluxes are listed for each ion in the same format as the results for the adopted UV fluxes. The best-fit value of the formation radius for each complex, derived using the adopted \tlusty\/ UV fluxes, is indicated in bold to facilitate comparison. }
\label{tab:Rfir_values1}
\end{table} 

\begin{table}
  \caption{Formation radii based on the {\it hegauss} fitting of the cycle 19 data }
\centering
\def\arraystretch{1.4}
\begin{tabular}{cccc}
  \hline
  S {\sc xv} & min & best & max \\
  \hline
  $0.67*H_\nu$ & 1.32 & 2.04 & unc. \\
  $H_\nu$ & 1.56 & {\bf 2.46} & unc. \\
  $1.5*H_\nu$ & 1.88 & 2.99 & unc. \\ 
  \hline
  Si {\sc xiii} &  &  &  \\
  \hline
  $0.67*H_\nu$ & 1.92 & 2.00 & 2.08 \\
  $H_\nu$ & 2.31 & {\bf 2.41} & 2.52 \\
  $1.5*H_\nu$ & 2.81 & 2.93 & 3.06 \\
  \hline
  Mg {\sc xi} &  &  &  \\
  \hline
  $0.67*H_\nu$ & 2.12 & 2.21 & 2.29 \\
  $H_\nu$ & 2.56 & {\bf 2.67} & 2.78 \\
  $1.5*H_\nu$ & 3.12 & 3.25 & 3.38 \\
  \hline
\end{tabular}
\\
{See notes for Table \ref{tab:Rfir_values1}. The upper confidence limits on \Rfir\/ are unconstrained for S {\sc xv}.}
\label{tab:Rfir_values19}
\end{table} 

% DIVIDING LINE BETWEEN hegauss and hewind

\begin{table*}
  \caption{Best-fit {\it hewind} parameters for S {\sc xv}, Si {\sc xiii}, and Mg {\sc xi}}   \vspace{1mm}
\centering
\def\arraystretch{1.4}
\begin{tabular}{cccc}
  \hline 
  Cycle 1 & S {\sc xv} & Si {\sc xiii} & Mg {\sc xi} \\
  \hline 
  $\tau_*$ & $0.11_{-0.11}^{+0.52}$ & $0.49_{-0.14}^{+0.17}$ & $0.92_{-0.18}^{+0.20}$ \\
  $R_0$ (\Rstar) & $1.43_{-0.15}^{+0.18}$ & $1.44_{-0.05}^{+0.06}$ & $1.49_{-0.06}^{+0.07}$ \\
  $\mathcal{G}$ & $0.56_{-0.24}^{+0.37}$ & $0.89_{-0.09}^{+0.10}$ & $0.59_{-0.05}^{+0.07}$ \\
  {\it hewind} norm (photons cm$^{-2}$ s$^{-1}$) & $1.84_{-0.20}^{+0.21}\times 10^{-5}$ & $9.14_{-0.28}^{+0.27}\times 10^{-5}$ & $1.52_{-0.04}^{+0.05}\times 10^{-4}$ \\
  {\it pow} norm (photons keV$^{-1}$ cm$^{-2}$ s$^{-1}$ at 1 keV) & $7.33_{-0.66}^{+0.70}\times 10^{-4}$ & $1.12_{-0.06}^{+0.06}\times 10^{-3}$ & $1.84_{-0.08}^{+0.08}\times 10^{-3}$ \\

  \hline
  Cycle 19 & S {\sc xv} & Si {\sc xiii} & Mg {\sc xi} \\
  \hline 
  $\tau_*$ & $0.26_{-0.12}^{+0.14}$ & $0.84_{-0.07}^{+0.08}$ & $0.96_{-0.09}^{+0.09}$ \\
  $R_0$ (\Rstar) & $1.49_{-0.07}^{+0.07}$ & $1.57_{-0.03}^{+0.03}$ & $1.71_{-0.03}^{+0.04}$ \\
  $\mathcal{G}$ & $0.77_{-0.10}^{+0.11}$ & $0.93_{-0.03}^{+0.03}$ & $0.66_{-0.03}^{+0.03}$ \\
  {\it hewind} norm (photons cm$^{-2}$ s$^{-1}$) & $2.15_{-0.07}^{+0.07}\times 10^{-5}$ & $1.06_{-0.01}^{+0.01}\times 10^{-4}$ & $1.74_{-0.02}^{+0.02}\times 10^{-4}$ \\
  {\it pow} norm (photons keV$^{-1}$ cm$^{-2}$ s$^{-1}$ at 1 keV) & $8.97_{-0.22}^{+0.23}\times 10^{-4}$ & $1.38_{-0.02}^{+0.02}\times 10^{-3}$ & $2.17_{-0.03}^{+0.03}\times 10^{-3}$ \\
 \hline
\label{tab:hewind}
\end{tabular}
\end{table*}

\clearpage 

\clearpage 

\section{Discussion}

The newer and higher signal-to-noise cycle 19 \chandra\/ spectra of \zpup\/ show unambiguously that the hottest plasma contributing to this O star's X-ray emission is well above the photosphere and in the outflowing wind. Notably, the S {\sc xv} characteristic formation radius is found at $\Rfir = 2.46$ \Rstar\/ (with 68 percent statistical lower confidence limit of $1.56$ \Rstar), whereas the formation radius is consistent with a location arbitrarily close to the photosphere in the cycle 1 data (with 68 percent statistical upper confidence limit of $1.27$ \Rstar)\footnote{While the cycle 1 {\it hegauss} results for S {\sc xv} are formally consistent with a formation location very close to the photosphere 
%-- even with the inclusion of satellite lines in the modeling -- 
they contradict the expectations of the wind-shock scenario with only slightly more than 1 sigma significance. Furthermore, the lines in this complex are Doppler broadened to an extent consistent with the high velocity of the wind, just as the other lines in the spectrum are.}. In the context of the {\it hewind} model, the onset radius for S {\sc xv} in the cycle 19 data is $\Ro = 1.49 \pm 0.07$ \Rstar, with an assumed density-squared weighted distribution above that; a result that is fully consistent with the expectations of LDI-generated embedded wind shocks\footnote{It should be noted that some more recent numerical simulations show wind structure forming at radii below the canonical $r \approx 1.5$ \Rstar, although it is not clear that that structure would be predicted to generate X-ray emission \citep{Sundqvist2018}.} and with no need to posit the existence of any hot plasma very close to the photosphere. Indeed, in the context of the {\it hewind} model, even the cycle 1 results for S {\sc xv} are consistent with the EWS scenario. Fig.\ \ref{fig:Ro_probs} shows that an onset radius, \Ro, above $r \approx 1.3$ \Rstar\/ is strongly favored, with a relatively broad probability distribution extending well above $r = 1.5$ \Rstar. 

Results from modeling the other two line complexes in both datasets are similar to the cycle 19 results for S {\sc xv}. Overall the characteristic radii of formation for Si {\sc xiii} and Mg {\sc xi} in both cycles range from $R_{\rm f/i}/\Rstar = 1.9$ to $R_{\rm f/i}/\Rstar = 2.7$ based on the {\it hegauss} best fit results. The onset radii, \Ro, derived from the {\it hewind} modeling, range from $\Ro/\Rstar = 1.49$ to $1.71$ in the cycle 19 data and slightly less ($\Ro/\Rstar = 1.43$ to $1.49$) in the cycle 1 data. 

We do not find evidence for the trend in the characteristic line-formation radius with atomic number (with the exception of S {\sc xv} in cycle 1) that is often found in Gaussian fitting and analysis of these complexes in OB stars \citep{wc2007}. This seems to be largely due to accounting for satellite line contamination in the fitting we present here. The trend we do find in the onset radius of the {\it hewind} model -- an anti-correlation between \Ro\/ and atomic number -- is small but statistically significant (at least in the cycle 19 data -- see Fig.\ \ref{fig:Ro_probs}).  In individual post-shock cooling zones one would indeed expect the hottest plasma and the emission from the higher ionization stages to be closer to the shock front and the star's surface than the somewhat cooler gas and its lower ionization stages. This, rather than any large-scale trend in shock strength with local wind velocity or distance from the star's surface, likely explains the very modest \Ro\/ trend we find, although we note that the recently developed {\it variable boundary conditions} EWS model \citep{Gunderson2022} may also be relevant here.

The particular line complexes we analyze in this paper are formed in the hottest plasma in the wind of \zpup. However, these line complexes do have some contribution from the less-hot plasma ($T =$ few $10^6$ K) that the LDI primarily produces and which is measured on O stars \citep[see fig.\/ 3 in][]{Cohen2021,Cohen2014a}. The $\mathcal{G} = (f+i)/r$ line ratios we find in the fitting are consistent with the expectations of the atomic physics \citep{Porquet2001, Foster2012} given the known shock-heated plasma temperature distribution in \zpup. All the results presented in this paper are consistent with the embedded wind shock scenario for O star wind X-ray production, including the onset radius of X-ray emission being well out in the wind flow. Although we do not address the short-term X-ray variability, linked to optical variability, seen in the cycle 19 \chandra\/ data \cite[1.78 day periodicity;][]{Nichols2021}, it is quite conceivable that it too is consistent with the EWS scenario, as photospheric variability can drive shock activity in the wind \citep{fpp1997}.

The cycle 19 data provide improved signal-to-noise compared to the cycle 1 data and tighter constraints on model parameters (see, e.g., the narrower probability distributions in the lower panel of Fig.\ \ref{fig:Ro_probs} and the tighter confidence limits on \Rfir\/ for Si {\sc xiii} and Mg {\sc xi} in Fig.\ \ref{fig:projRfir}), and they also indicate that there have been some changes to the hot plasma properties in the wind of \zpup\/ in the 18-year interval between the two sets of observations. As detailed by \citet{Cohen2020}, the line shapes have changed and the line fluxes have increased between the two epochs in a way consistent with an increase in the wind mass-loss rate between the two observing campaigns. In the analysis we present here, these changes are manifest in the overall line-complex and continuum fluxes, which increased by 10 to 15 per cent between cycle 1 and cycle 19, by an increase in the degree of wind absorption (\taustar\/ in the {\it hewind} modeling), and also by an increase in the line widths ($\sigma$ in the {\it hegauss} modeling). These increased line widths might be due to changes in the kinematics of the shocked wind plasma and/or they might be due to changes in the location, with the newer data being consistent with X-ray production somewhat farther out in the wind flow (a modest result in the {\it hewind} modeling -- see Fig.\ \ref{fig:Ro_probs} -- and a more significant result for S {\sc xv} and Mg {\sc xi} in the {\it hegauss} modeling -- see Fig.\ \ref{fig:projRfir}). 

For both the interpretation of the {\it hegauss} fitting results and for the {\it hewind} fitting itself, the adopted photospheric UV flux values are important. If we conservatively assume a UV flux uncertainty, to account for systematics in the atmosphere modeling as well as in the stellar parameters, of $\pm 50$ per cent, we find uncertainties in the formation radius in the {\it hegauss} modeling of roughly 20 to 30 percent (in terms of relative distance from the photosphere -- see tables \ref{tab:Rfir_values1} and \ref{tab:Rfir_values19}). The UV radiation driving the S {\sc xv} and Si {\sc xiii} line ratios is shortward of the hydrogen Lyman edge and therefore both quite sensitive to the temperature gradient in the model atmosphere and also unobservable in \zpup\/ or any other O star, hence there will likely always be a fair amount of uncertainty in the modeling of these line complexes. 

Despite the uncertainty in the UV fluxes and resultant photoexcitation rates, the primary results presented here are secure: the X-ray emitting plasma is located well out in the wind flow ($ > 1.4$ \Rstar), with no hot plasma close to the photosphere required to explain the observed features; and there was a brightening of the X-rays, increased line widths, and an increase in the radial location of the X-ray emitting plasma between the data taken in 2000 and those taken in 2018-19. Finally, the inclusion of DR satellite lines in the He-complex modeling is tractable -- especially if a hot plasma temperature distribution is assumed -- and leads to lower He-like line fluxes and modest alteration to the other fitted model parameters and significant changes to the behavior of $\mathcal{R} \equiv f/i$ at its two asymptotic limits. We stress again that all the He-like modeling results presented here are corrected for the effects of the satellite lines by the inclusion of the latter in the spectral model fitting, as detailed in \S3 and in the appendix. 

The next steps in modeling He-like complexes in OB stars should include refinement of the DR satellite atomic model -- using the {\it Flexible Atomic Code} \citep{Gu2008} for updated line strengths and including higher $n$ transitions. Another improvement to the modeling would treat the photospheric UV transport through the wind in detail. This could make a noticeable difference in the interpretation of the Mg {\sc xi} diagnostic, as the photoexcitation wavelength is coincident with a strong O {\sc vi} wind line and redistribution associated with the resultant scattering of the UV radiation could decrease the UV mean intensity at the location of the X-ray emitting plasma. Evaluating this will require detailed modeling with a complex non-LTE wind code, and it could reduce or even eliminate the observed trend of \Ro\/ with atomic number and plasma temperature. 

\section*{Acknowledgements}

The scientific results in this article are based on data retrieved from the \chandra\/ data archive. DHC acknowledges support from the {\it Chandra} Guest Observer Program via award AR1-22001A to Swarthmore College and support from the Provost's Office of Swarthmore College. AMO acknowledges support from the Provost's Office of Swarthmore College via the Eugene M.\ Lang Summer Research Fellowship. MAL and ADU acknowledge support from NASA awards 18-2ADAP18-0149 and 80GSFC21M0002 and from the {\it Chandra} Guest Observer Program via award AR1-22001Z.We thank Jiaming Wang for help with some of the figures in the paper. And we thank the anonymous referee for suggestions that improved the paper.  

\section*{Data Availability}

The X-ray spectral data underlying this article are available in the
Chandra Data Archive at \url{https://cxc.cfa.harvard.edu/cda/}, as discussed in \S2, with a detailed observing log available in table 1 of \citet{Cohen2020}.

%\clearpage

\bibliographystyle{mnras}
\bibliography{zpup}

\begin{thebibliography}{}
\makeatletter
\relax
\def\mn@urlcharsother{\let\do\@makeother \do\$\do\&\do\#\do\^\do\_\do\%\do\~}
\def\mn@doi{\begingroup\mn@urlcharsother \@ifnextchar [ {\mn@doi@}
  {\mn@doi@[]}}
\def\mn@doi@[#1]#2{\def\@tempa{#1}\ifx\@tempa\@empty \href
  {http://dx.doi.org/#2} {doi:#2}\else \href {http://dx.doi.org/#2} {#1}\fi
  \endgroup}
\def\mn@eprint#1#2{\mn@eprint@#1:#2::\@nil}
\def\mn@eprint@arXiv#1{\href {http://arxiv.org/abs/#1} {{\tt arXiv:#1}}}
\def\mn@eprint@dblp#1{\href {http://dblp.uni-trier.de/rec/bibtex/#1.xml}
  {dblp:#1}}
\def\mn@eprint@#1:#2:#3:#4\@nil{\def\@tempa {#1}\def\@tempb {#2}\def\@tempc
  {#3}\ifx \@tempc \@empty \let \@tempc \@tempb \let \@tempb \@tempa \fi \ifx
  \@tempb \@empty \def\@tempb {arXiv}\fi \@ifundefined
  {mn@eprint@\@tempb}{\@tempb:\@tempc}{\expandafter \expandafter \csname
  mn@eprint@\@tempb\endcsname \expandafter{\@tempc}}}

\bibitem[\protect\citeauthoryear{{Blumenthal}, {Drake}  \&
  {Tucker}}{{Blumenthal} et~al.}{1972}]{bdt1972}
{Blumenthal} G.~R.,  {Drake} G.~W.~F.,   {Tucker} W.~H.,  1972, \mn@doi [\apj]
  {10.1086/151340}, \href
  {https://ui.adsabs.harvard.edu/abs/1972ApJ...172..205B} {172, 205}

\bibitem[\protect\citeauthoryear{{Bouret}, {Hillier}, {Lanz}  \&
  {Fullerton}}{{Bouret} et~al.}{2012}]{Bouret2012}
{Bouret} J.~C.,  {Hillier} D.~J.,  {Lanz} T.,   {Fullerton} A.~W.,  2012,
  \mn@doi [\aap] {10.1051/0004-6361/201118594}, \href
  {https://ui.adsabs.harvard.edu/abs/2012A&A...544A..67B} {544, A67}

\bibitem[\protect\citeauthoryear{{Bryans}, {Badnell}, {Gorczyca}, {Laming},
  {Mitthumsiri}  \& {Savin}}{{Bryans} et~al.}{2006}]{Bryans2006}
{Bryans} P.,  {Badnell} N.~R.,  {Gorczyca} T.~W.,  {Laming} J.~M.,
  {Mitthumsiri} W.,   {Savin} D.~W.,  2006, \mn@doi [\apjs] {10.1086/507629},
  \href {https://ui.adsabs.harvard.edu/abs/2006ApJS..167..343B} {167, 343}

\bibitem[\protect\citeauthoryear{{Bryans}, {Landi}  \& {Savin}}{{Bryans}
  et~al.}{2009}]{Bryans2009}
{Bryans} P.,  {Landi} E.,   {Savin} D.~W.,  2009, \mn@doi [\apj]
  {10.1088/0004-637X/691/2/1540}, \href
  {https://ui.adsabs.harvard.edu/abs/2009ApJ...691.1540B} {691, 1540}

\bibitem[\protect\citeauthoryear{{Cash}}{{Cash}}{1979}]{Cash1979}
{Cash} W.,  1979, \mn@doi [\apj] {10.1086/156922}, \href
  {https://ui.adsabs.harvard.edu/abs/1979ApJ...228..939C} {228, 939}

\bibitem[\protect\citeauthoryear{{Cassinelli}, {Miller}, {Waldron},
  {MacFarlane}  \& {Cohen}}{{Cassinelli} et~al.}{2001}]{Cassinelli2001}
{Cassinelli} J.~P.,  {Miller} N.~A.,  {Waldron} W.~L.,  {MacFarlane} J.~J.,
  {Cohen} D.~H.,  2001, \mn@doi [\apjl] {10.1086/320916}, \href
  {https://ui.adsabs.harvard.edu/abs/2001ApJ...554L..55C} {554, L55}

\bibitem[\protect\citeauthoryear{{Cohen}, {Wollman}, {Leutenegger},
  {Sundqvist}, {Fullerton}, {Zsarg{\'o}}  \& {Owocki}}{{Cohen}
  et~al.}{2014a}]{Cohen2014b}
{Cohen} D.~H.,  {Wollman} E.~E.,  {Leutenegger} M.~A.,  {Sundqvist} J.~O.,
  {Fullerton} A.~W.,  {Zsarg{\'o}} J.,   {Owocki} S.~P.,  2014a, \mn@doi
  [\mnras] {10.1093/mnras/stu008}, \href
  {http://adsabs.harvard.edu/abs/2014MNRAS.439..908C} {439, 908}

\bibitem[\protect\citeauthoryear{{Cohen}, {Li}, {Gayley}, {Owocki},
  {Sundqvist}, {Petit}  \& {Leutenegger}}{{Cohen} et~al.}{2014b}]{Cohen2014a}
{Cohen} D.~H.,  {Li} Z.,  {Gayley} K.~G.,  {Owocki} S.~P.,  {Sundqvist} J.~O.,
  {Petit} V.,   {Leutenegger} M.~A.,  2014b, \mn@doi [\mnras]
  {10.1093/mnras/stu1661}, \href
  {https://ui.adsabs.harvard.edu/abs/2014MNRAS.444.3729C} {444, 3729}

\bibitem[\protect\citeauthoryear{{Cohen}, {Wang}, {Petit}, {Leutenegger},
  {Dakir}, {Mayhue}  \& {David-Uraz}}{{Cohen} et~al.}{2020}]{Cohen2020}
{Cohen} D.~H.,  {Wang} J.,  {Petit} V.,  {Leutenegger} M.~A.,  {Dakir} L.,
  {Mayhue} C.,   {David-Uraz} A.,  2020, \mn@doi [\mnras]
  {10.1093/mnras/staa3124}, \href
  {https://ui.adsabs.harvard.edu/abs/2020MNRAS.499.6044C} {499, 6044}

\bibitem[\protect\citeauthoryear{{Cohen}, {Parts}, {Doskoch}, {Wang}, {Petit},
  {Leutenegger}  \& {Gagn{\'e}}}{{Cohen} et~al.}{2021}]{Cohen2021}
{Cohen} D.~H.,  {Parts} W.,  {Doskoch} G.~M.,  {Wang} J.,  {Petit} V.,
  {Leutenegger} M.~A.,   {Gagn{\'e}} M.,  2021, \mn@doi [\mnras]
  {10.1093/mnras/stab270}, \href
  {https://ui.adsabs.harvard.edu/abs/2021MNRAS.503..715C} {503, 715}

\bibitem[\protect\citeauthoryear{{Dorman} \& {Arnaud}}{{Dorman} \&
  {Arnaud}}{2001}]{Dorman2001}
{Dorman} B.,  {Arnaud} K.~A.,  2001, in {Harnden} F.~R. J.,  {Primini} F.~A.,
  {Payne} H.~E.,  eds,  Astronomical Society of the Pacific Conference Series
  Vol. 238, Astronomical Data Analysis Software and Systems X. p.~415

\bibitem[\protect\citeauthoryear{{Feldmeier}, {Puls}  \&
  {Pauldrach}}{{Feldmeier} et~al.}{1997}]{fpp1997}
{Feldmeier} A.,  {Puls} J.,   {Pauldrach} A.~W.~A.,  1997, \aap, \href
  {https://ui.adsabs.harvard.edu/abs/1997A&A...322..878F} {322, 878}

\bibitem[\protect\citeauthoryear{{Foster} \& {Heuer}}{{Foster} \&
  {Heuer}}{2020}]{Foster2020}
{Foster} A.~R.,  {Heuer} K.,  2020, \mn@doi [Atoms] {10.3390/atoms8030049},
  \href {https://ui.adsabs.harvard.edu/abs/2020Atoms...8...49F} {8, 49}

\bibitem[\protect\citeauthoryear{{Foster}, {Ji}, {Smith}  \&
  {Brickhouse}}{{Foster} et~al.}{2012}]{Foster2012}
{Foster} A.~R.,  {Ji} L.,  {Smith} R.~K.,   {Brickhouse} N.~S.,  2012, \mn@doi
  [\apj] {10.1088/0004-637X/756/2/128}, \href
  {https://ui.adsabs.harvard.edu/abs/2012ApJ...756..128F} {756, 128}

\bibitem[\protect\citeauthoryear{{Gabriel} \& {Jordan}}{{Gabriel} \&
  {Jordan}}{1969}]{gj1969}
{Gabriel} A.~H.,  {Jordan} C.,  1969, \mn@doi [\mnras]
  {10.1093/mnras/145.2.241}, \href
  {https://ui.adsabs.harvard.edu/abs/1969MNRAS.145..241G} {145, 241}

\bibitem[\protect\citeauthoryear{{Gaia Collaboration} et~al.,}{{Gaia
  Collaboration} et~al.}{2021}]{GAIA2021}
{Gaia Collaboration} et~al., 2021, \mn@doi [\aap]
  {10.1051/0004-6361/202039657}, \href
  {https://ui.adsabs.harvard.edu/abs/2021A&A...649A...1G} {649, A1}

\bibitem[\protect\citeauthoryear{{Gu}}{{Gu}}{2008}]{Gu2008}
{Gu} M.~F.,  2008, \mn@doi [Canadian Journal of Physics] {10.1139/P07-197},
  \href {https://ui.adsabs.harvard.edu/abs/2008CaJPh..86..675G} {86, 675}

\bibitem[\protect\citeauthoryear{{Gunderson}, {Gayley}, {Pradhan},
  {Huenemoerder}  \& {Miller}}{{Gunderson} et~al.}{2022}]{Gunderson2022}
{Gunderson} S.~J.,  {Gayley} K.~G.,  {Pradhan} P.,  {Huenemoerder} D.~P.,
  {Miller} N.~A.,  2022, \mn@doi [\mnras] {10.1093/mnras/stac552}, \href
  {https://ui.adsabs.harvard.edu/abs/2022MNRAS.tmp..549G} {}

\bibitem[\protect\citeauthoryear{{Hainich}, {Ramachandran}, {Shenar}, {Sander},
  {Todt}, {Gruner}, {Oskinova}  \& {Hamann}}{{Hainich}
  et~al.}{2019}]{Hainich2019}
{Hainich} R.,  {Ramachandran} V.,  {Shenar} T.,  {Sander} A.~A.~C.,  {Todt} H.,
   {Gruner} D.,  {Oskinova} L.~M.,   {Hamann} W.~R.,  2019, \mn@doi [\aap]
  {10.1051/0004-6361/201833787}, \href
  {https://ui.adsabs.harvard.edu/abs/2019A&A...621A..85H} {621, A85}

\bibitem[\protect\citeauthoryear{{Haser}}{{Haser}}{1995}]{Haser1995}
{Haser} S.~M.,  1995, PhD thesis, Universit\"{a}ts-Sternwarte der
  Ludwig-Maximillian Universit\"{a}t, M\"{u}nchen

\bibitem[\protect\citeauthoryear{{Howarth} \& {van Leeuwen}}{{Howarth} \& {van
  Leeuwen}}{2019}]{Howarth2019}
{Howarth} I.~D.,  {van Leeuwen} F.,  2019, \mn@doi [\mnras]
  {10.1093/mnras/stz291}, \href
  {https://ui.adsabs.harvard.edu/abs/2019MNRAS.484.5350H} {484, 5350}

\bibitem[\protect\citeauthoryear{{Huenemoerder} et~al.,}{{Huenemoerder}
  et~al.}{2020}]{Huenemoerder2020}
{Huenemoerder} D.~P.,  et~al., 2020, \mn@doi [\apj] {10.3847/1538-4357/ab8005},
  \href {https://ui.adsabs.harvard.edu/abs/2020ApJ...893...52H} {893, 52}

\bibitem[\protect\citeauthoryear{{Kahn}, {Leutenegger}, {Cottam}, {Rauw},
  {Vreux}, {den Boggende}, {Mewe}  \& {G{\"u}del}}{{Kahn}
  et~al.}{2001}]{Kahn2001}
{Kahn} S.~M.,  {Leutenegger} M.~A.,  {Cottam} J.,  {Rauw} G.,  {Vreux} J.~M.,
  {den Boggende} A.~J.~F.,  {Mewe} R.,   {G{\"u}del} M.,  2001, \mn@doi [\aap]
  {10.1051/0004-6361:20000093}, \href
  {https://ui.adsabs.harvard.edu/abs/2001A&A...365L.312K} {365, L312}

\bibitem[\protect\citeauthoryear{{Lanz} \& {Hubeny}}{{Lanz} \&
  {Hubeny}}{2003}]{Lanz2003}
{Lanz} T.,  {Hubeny} I.,  2003, \mn@doi [\apjs] {10.1086/374373}, \href
  {https://ui.adsabs.harvard.edu/abs/2003ApJS..146..417L} {146, 417}

\bibitem[\protect\citeauthoryear{{Leutenegger}, {Paerels}, {Kahn}  \&
  {Cohen}}{{Leutenegger} et~al.}{2006}]{Leutenegger2006}
{Leutenegger} M.~A.,  {Paerels} F. B.~S.,  {Kahn} S.~M.,   {Cohen} D.~H.,
  2006, \mn@doi [\apj] {10.1086/507147}, \href
  {https://ui.adsabs.harvard.edu/abs/2006ApJ...650.1096L} {650, 1096}

\bibitem[\protect\citeauthoryear{{Martins} et~al.,}{{Martins}
  et~al.}{2015}]{Martins2015}
{Martins} F.,  et~al., 2015, \mn@doi [\aap] {10.1051/0004-6361/201425173},
  \href {https://ui.adsabs.harvard.edu/abs/2015A&A...575A..34M} {575, A34}

\bibitem[\protect\citeauthoryear{{Massa} et~al.,}{{Massa}
  et~al.}{1995}]{Massa1995}
{Massa} D.,  et~al., 1995, \mn@doi [\apjl] {10.1086/309707}, \href
  {https://ui.adsabs.harvard.edu/abs/1995ApJ...452L..53M} {452, L53}

\bibitem[\protect\citeauthoryear{{Nichols} et~al.,}{{Nichols}
  et~al.}{2021}]{Nichols2021}
{Nichols} J.~S.,  et~al., 2021, \mn@doi [\apj] {10.3847/1538-4357/abca3a},
  \href {https://ui.adsabs.harvard.edu/abs/2021ApJ...906...89N} {906, 89}

\bibitem[\protect\citeauthoryear{{Nousek} \& {Shue}}{{Nousek} \&
  {Shue}}{1989}]{Nousek1989}
{Nousek} J.~A.,  {Shue} D.~R.,  1989, \mn@doi [\apj] {10.1086/167676}, \href
  {https://ui.adsabs.harvard.edu/abs/1989ApJ...342.1207N} {342, 1207}

\bibitem[\protect\citeauthoryear{{Owocki} \& {Cohen}}{{Owocki} \&
  {Cohen}}{2001}]{OC2001}
{Owocki} S.~P.,  {Cohen} D.~H.,  2001, \mn@doi [\apj] {10.1086/322413}, \href
  {https://ui.adsabs.harvard.edu/abs/2001ApJ...559.1108O} {559, 1108}

\bibitem[\protect\citeauthoryear{{Owocki}, {Castor}  \& {Rybicki}}{{Owocki}
  et~al.}{1988}]{ocr1988}
{Owocki} S.~P.,  {Castor} J.~I.,   {Rybicki} G.~B.,  1988, \mn@doi [\apj]
  {10.1086/166977}, \href
  {https://ui.adsabs.harvard.edu/abs/1988ApJ...335..914O} {335, 914}

\bibitem[\protect\citeauthoryear{{Pauldrach}, {Vanbeveren}  \&
  {Hoffmann}}{{Pauldrach} et~al.}{2012}]{Pauldrach2012}
{Pauldrach} A.~W.~A.,  {Vanbeveren} D.,   {Hoffmann} T.~L.,  2012, \mn@doi
  [\aap] {10.1051/0004-6361/201117621}, \href
  {https://ui.adsabs.harvard.edu/abs/2012A&A...538A..75P} {538, A75}

\bibitem[\protect\citeauthoryear{{Porquet}, {Mewe}, {Dubau}, {Raassen}  \&
  {Kaastra}}{{Porquet} et~al.}{2001}]{Porquet2001}
{Porquet} D.,  {Mewe} R.,  {Dubau} J.,  {Raassen} A.~J.~J.,   {Kaastra} J.~S.,
  2001, \mn@doi [\aap] {10.1051/0004-6361:20010959}, \href
  {https://ui.adsabs.harvard.edu/abs/2001A&A...376.1113P} {376, 1113}

\bibitem[\protect\citeauthoryear{{Ramiaramanantsoa} et~al.,}{{Ramiaramanantsoa}
  et~al.}{2018}]{Tahina2018}
{Ramiaramanantsoa} T.,  et~al., 2018, \mn@doi [\mnras] {10.1093/mnras/stx2671},
  \href {https://ui.adsabs.harvard.edu/abs/2018MNRAS.473.5532R} {473, 5532}

\bibitem[\protect\citeauthoryear{{Runacres} \& {Owocki}}{{Runacres} \&
  {Owocki}}{2002}]{ro2002}
{Runacres} M.~C.,  {Owocki} S.~P.,  2002, \mn@doi [\aap]
  {10.1051/0004-6361:20011526}, \href
  {https://ui.adsabs.harvard.edu/abs/2002A&A...381.1015R} {381, 1015}

\bibitem[\protect\citeauthoryear{Sota, Apell{\'{a}}niz, Morrell, Barb{\'{a}},
  Walborn, Gamen, Arias  \& Alfaro}{Sota et~al.}{2014}]{Sota2014}
Sota A.,  Apell{\'{a}}niz J.~M.,  Morrell N.~I.,  Barb{\'{a}} R.~H.,  Walborn
  N.~R.,  Gamen R.~C.,  Arias J.~I.,   Alfaro E.~J.,  2014, \mn@doi [The
  Astrophysical Journal Supplement Series] {10.1088/0067-0049/211/1/10}, 211,
  10

\bibitem[\protect\citeauthoryear{{Sundqvist}, {Owocki}  \& {Puls}}{{Sundqvist}
  et~al.}{2018}]{Sundqvist2018}
{Sundqvist} J.~O.,  {Owocki} S.~P.,   {Puls} J.,  2018, \mn@doi [\aap]
  {10.1051/0004-6361/201731718}, \href
  {https://ui.adsabs.harvard.edu/abs/2018A&A...611A..17S} {611, A17}

\bibitem[\protect\citeauthoryear{{Waldron} \& {Cassinelli}}{{Waldron} \&
  {Cassinelli}}{2001}]{wc2001}
{Waldron} W.~L.,  {Cassinelli} J.~P.,  2001, \mn@doi [\apjl] {10.1086/318926},
  \href {https://ui.adsabs.harvard.edu/abs/2001ApJ...548L..45W} {548, L45}

\bibitem[\protect\citeauthoryear{{Waldron} \& {Cassinelli}}{{Waldron} \&
  {Cassinelli}}{2007}]{wc2007}
{Waldron} W.~L.,  {Cassinelli} J.~P.,  2007, \mn@doi [\apj] {10.1086/520919},
  \href {https://ui.adsabs.harvard.edu/abs/2007ApJ...668..456W} {668, 456}

\makeatother
\end{thebibliography}

%\clearpage 

\appendix
\section{Dielectronic satellite line modeling} \label{appendix:satellites}

As discussed in \S3 there are numerous dielectronic recombination (DR) satellite lines near, and potentially blended with, the He-like lines in X-ray spectra. These DR lines are generally weak, but there are many of them and given the wind Doppler broadening of emission lines in the \chandra\/ spectra of \zpup, the satellite lines in the aggregate contribute significantly to each He-like line. To account for this effect we compute detailed spectral models using \apec\/ \citep{Foster2012}, assuming a continuous plasma temperature distribution (differential emission measure, DEM) taken to be a power-law function with an index of $n= -2.3$ \citep{Cohen2021}. Note that the DR satellites form more readily at lower temperatures than do the He-like lines and thus this relatively steep DEM slope seen in O stars makes the satellite contamination a more significant factor than in many other types of sources, such as low-mass stars with coronal activity. 

\begin{figure}
\centering
    \includegraphics[angle=0,width=0.47\textwidth]{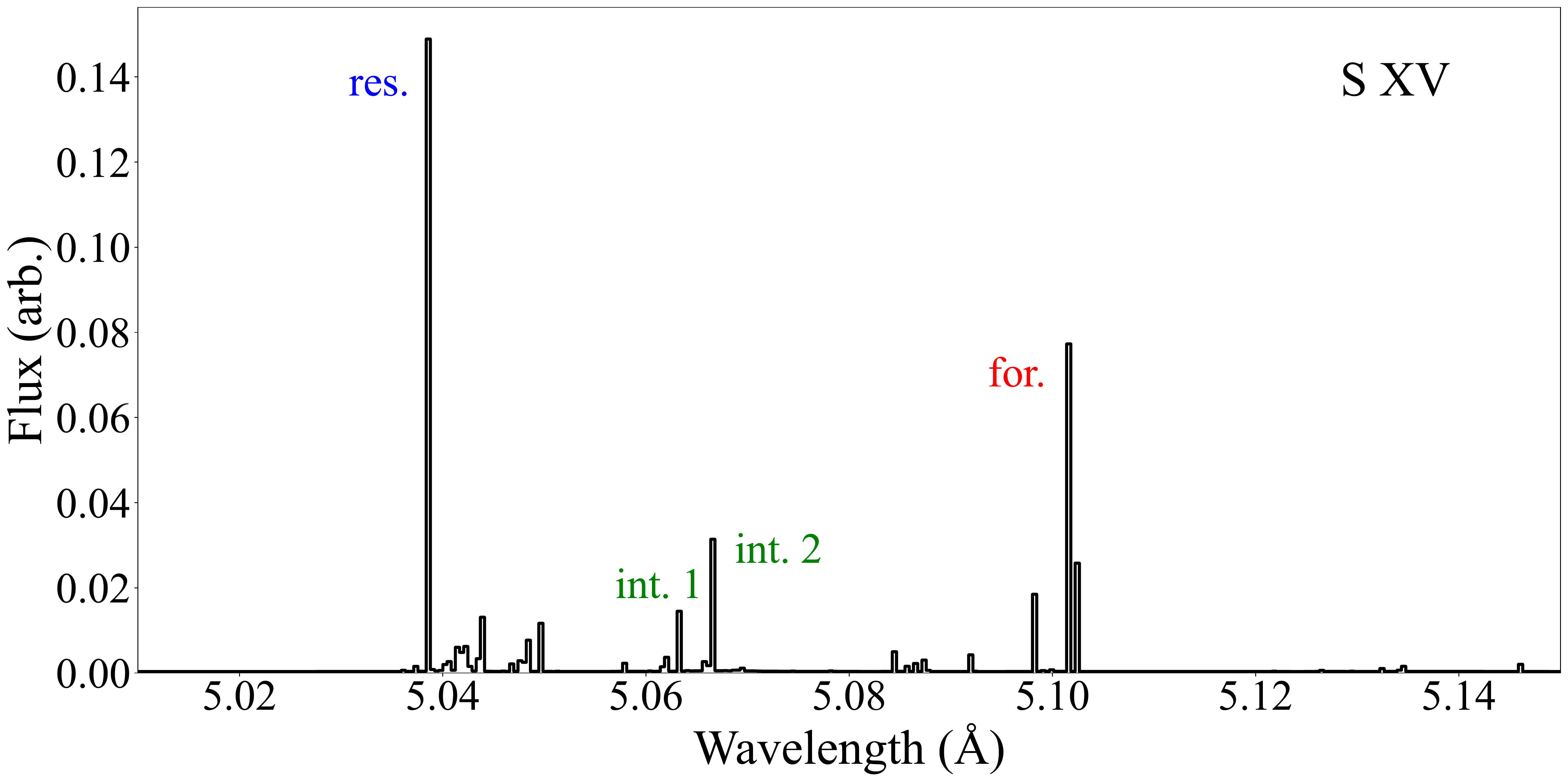}
    \includegraphics[angle=0,width=0.47\textwidth]{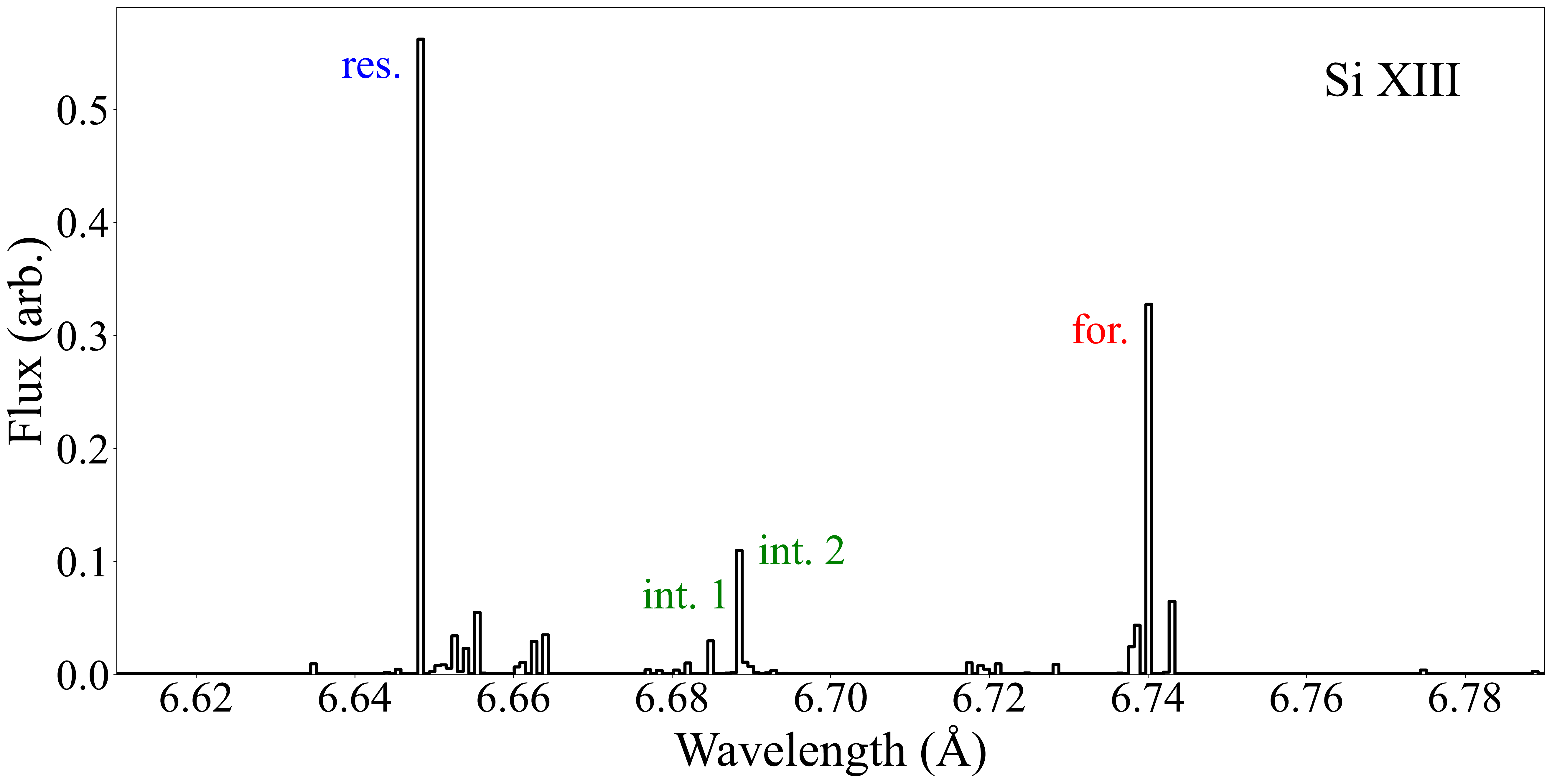}
    \includegraphics[angle=0,width=0.47\textwidth]{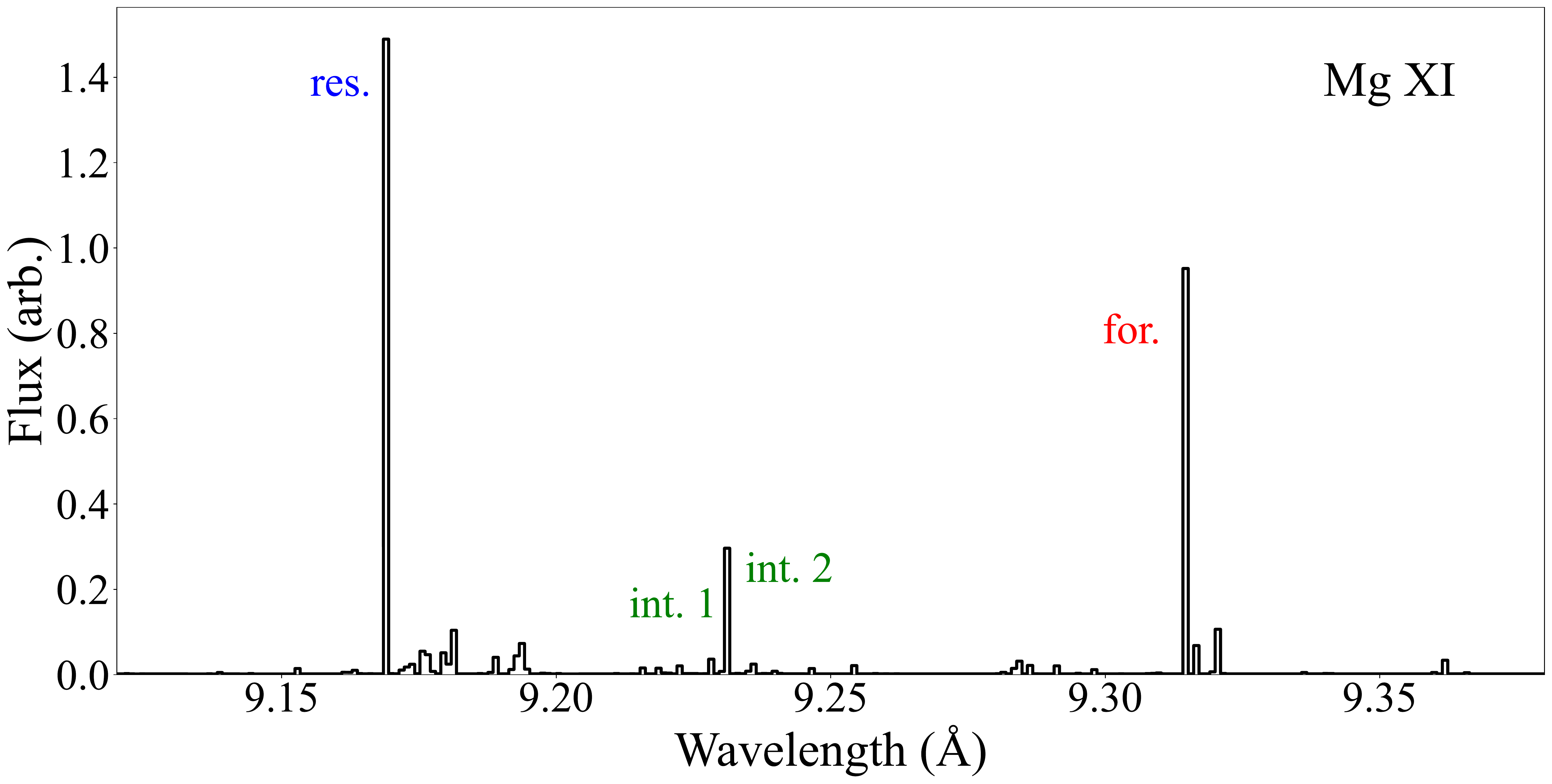}
    \caption{\apec\/ models of the He-like complexes (S {\sc xv}, Si {\sc xiii}, Mg {\sc xi}, top to bottom) and associated DR satellite lines shown at high resolution. The flux levels are arbitrary but self-consistent among the three ions and are specific to our assumed DEM model. The resonance line, two intercombination components, and forbidden line are labeled in each panel.}
	\label{fig:satellites_model}
\end{figure}

\begin{figure}
\centering
    \includegraphics[angle=0,width=0.47\textwidth]{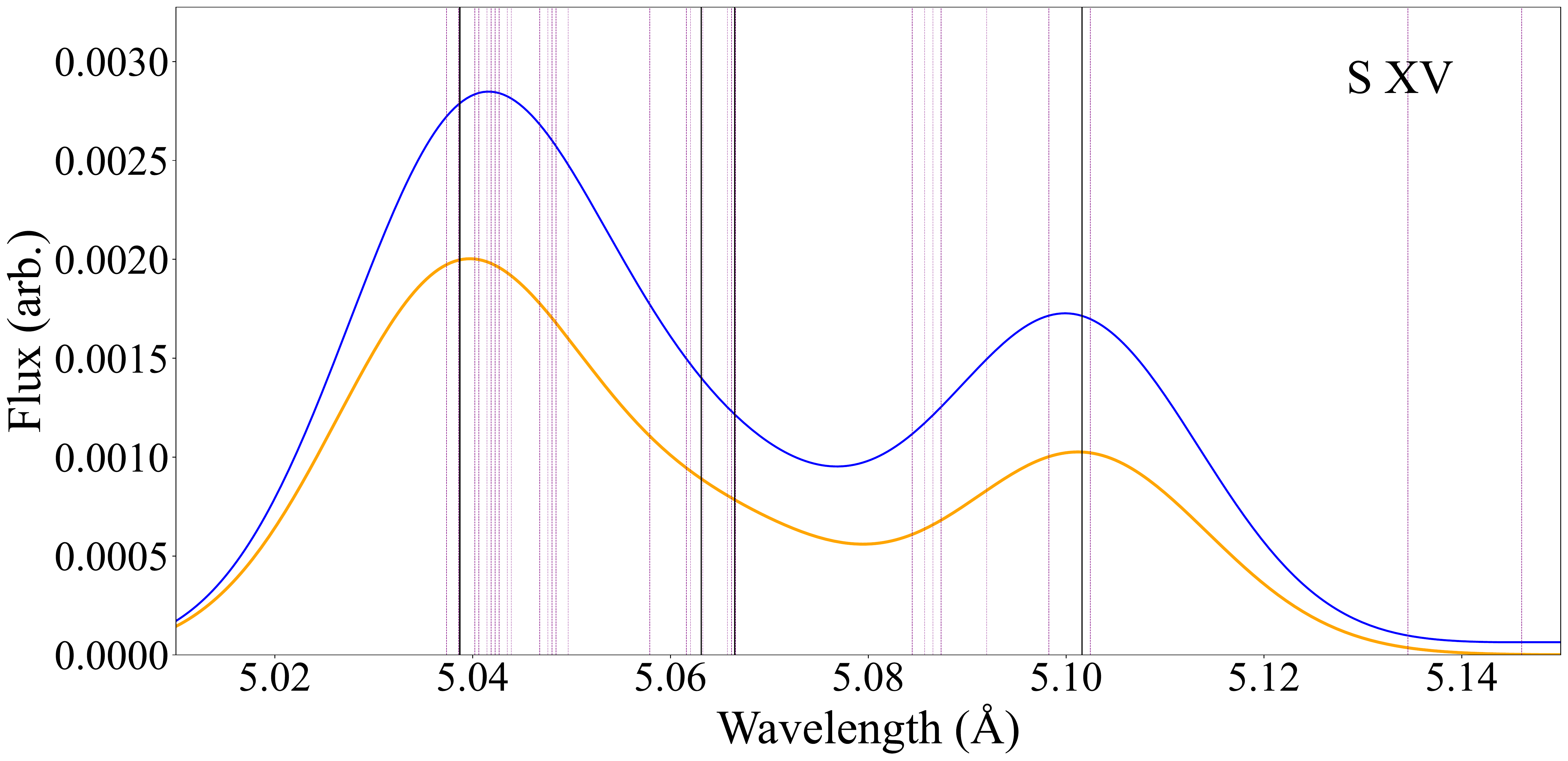}
    \includegraphics[angle=0,width=0.47\textwidth]{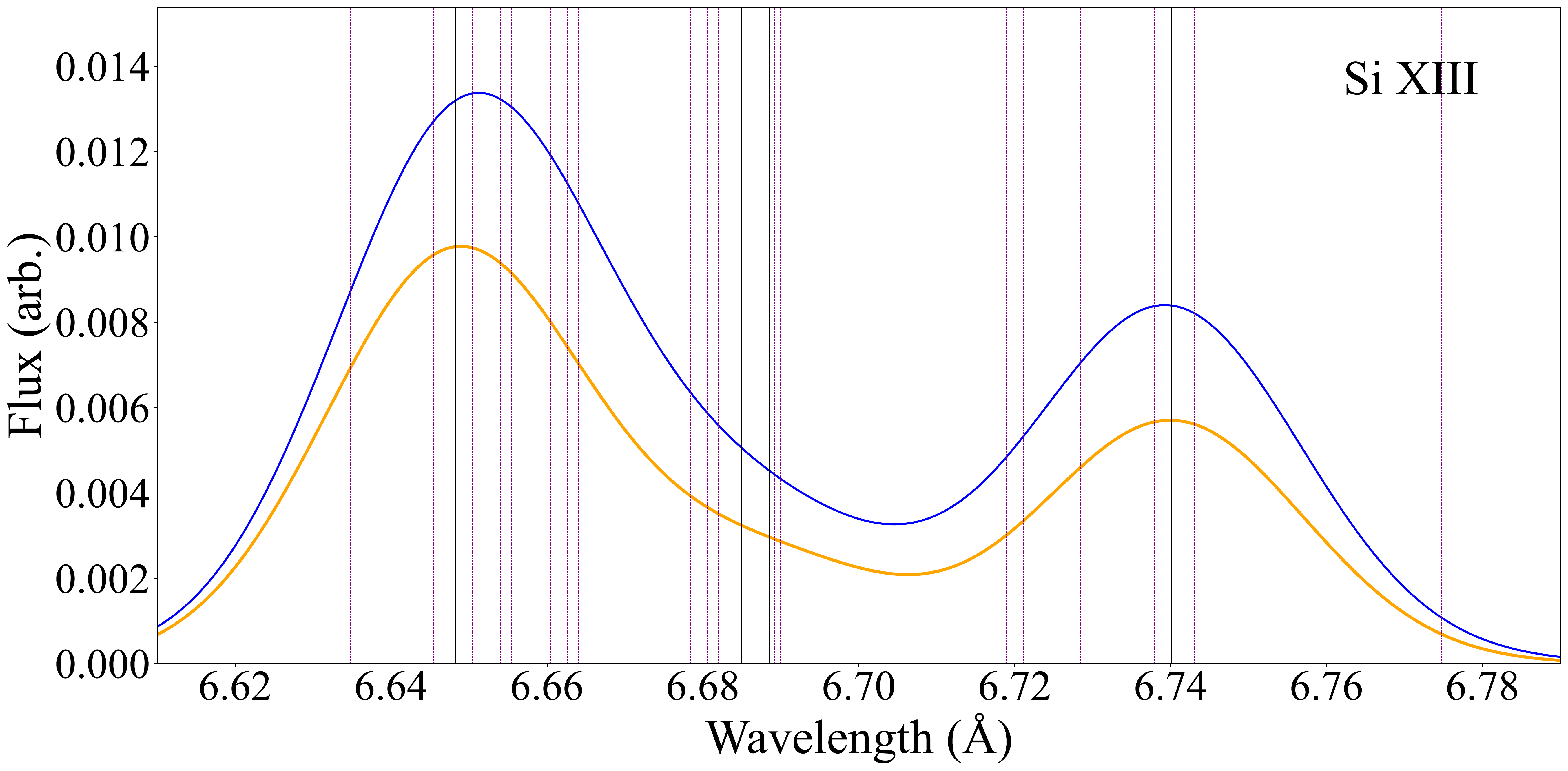}
    \includegraphics[angle=0,width=0.47\textwidth]{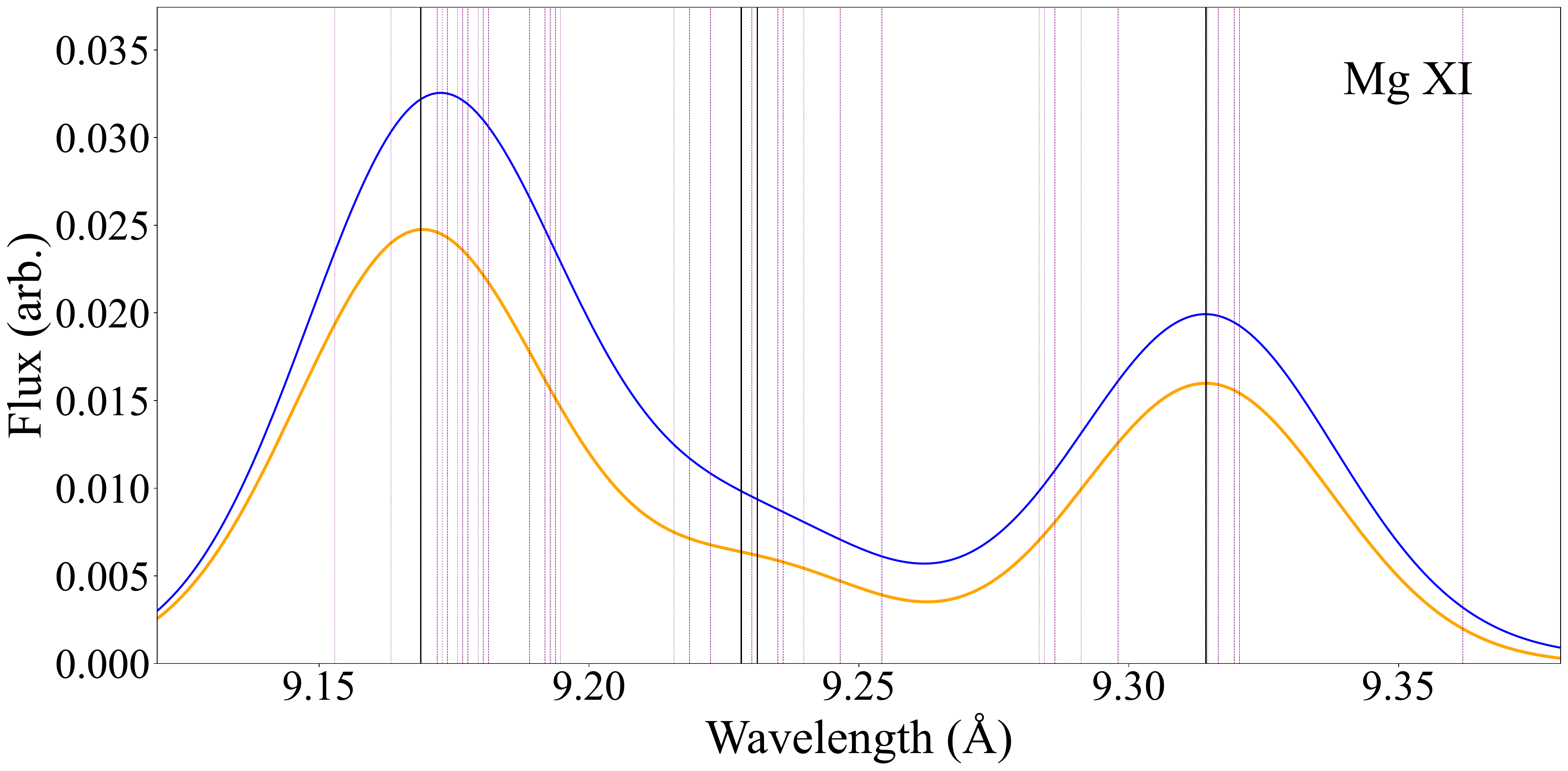}
    \caption{He-like complexes including DR satellite lines (blue), and the contribution just from the He-like lines (orange), are shown with Gaussian broadening ($\sigma = 750$ \kms). The wavelengths of the satellite lines included in the modeling -- both that shown here and the spectral model fitting of the data themselves -- are indicated by vertical dashed purple lines and those of the four He-like components by vertical black lines.}
	\label{fig:satellites_convolved}
\end{figure}

We show the \apec\/ spectral models in Fig.\ \ref{fig:satellites_model}, which confirm that in the aggregate, the satellite contribution\footnote{Some of the weak, contaminating lines seen in the \apec\/ models are not DR satellites but rather are due to other elements, including some high ionization iron lines and a few Ne {\sc x} lines in the Mg {\sc xi} complex and Mg {\sc xii} near the Si {\sc xiii} forbidden line.} is a non-negligible fraction of the forbidden and intercombination lines for each ion. We show the model spectra with Gaussian line broadening assuming a width of $\sigma = 750$ \kms, which is the typical line broadening seen in the \chandra\/ spectra of \zpup, in Fig.\ \ref{fig:satellites_convolved}. We found that the satellite contributions to the line complex can be accounted for with 29, 30, and 39 satellites for S {\sc xv}, Si {\sc xiii}, and Mg {\sc xi}, respectively. To incorporate these satellite lines into the model fitting reported on in this paper we added individual Gaussians (for the {\it hegauss} model) or individual wind-profile models (for the {\it hewind} model) for each of the satellites. We tied the satellite line profile parameters (width and shift in the case of the Gaussians and \taustar\/ and \Ro\/ for the wind profile models) to the corresponding parameters in {\it hegauss} or {\it hewind} and scaled the flux of each satellite line to that of the total flux of the He-like components. Note that this does not add any free parameters to either model. We find improvement to the fit statistic when we include the satellites in most cases (these fit statistic values are listed in Table \ref{tab:fit_stats}). 

\begin{table*}
\caption{Fit statistics with and without the inclusion of satellite lines in the data fitting 
}
\centering
\begin{tabular}{cccccc}
  \hline
  ion and observation & $N$ & ${\rm C}_{\rm sats}$ {\it hegauss} & ${\rm C}_{\rm no~sats}$ {\it hegauss}   & ${\rm C}_{\rm sats}$ {\it hewind} & ${\rm C}_{\rm no~sats}$ {\it hewind} \\
  \hline
  S {\sc xv} cycle 1 & 208 & 180.94 & 183.82 & 183.70 & 184.57 \\
    Si {\sc xiii} cycle 1 & 280 & 325.72 & 320.84 & 313.68 & 319.29 \\
    Mg {\sc xi} cycle 1 & 346 & 402.84 & 416.05 & 397.57 & 392.74 \\
     S {\sc xv} cycle 19 & 3468 & 3247.66 & 3249.89 & 3245.82 & 3251.32 \\
    Si {\sc xiii} cycle 19 & 5880 & 6473.01 & 6404.81 & 6427.04 & 6444.28 \\
    Mg {\sc xi} cycle 19 & 7644 & 8520.44 & 8540.31 & 8512.41 & 8516.06 \\

    \hline
\end{tabular}
\\
{
The second column lists the number of data points included in a given fit. The third and fourth columns list the best-fit {\it hegauss} model's C statistic with and without the inclusion of satellites. The last two columns list the best-fit {\it hewind} model's C statistic with and without the inclusion of satellites. For both models, the ``no satellites'' case involves re-fitting all the free model parameters of the He-like model but without any satellite lines included in the modeling. 
}
\label{tab:fit_stats}
\end{table*} 

%\clearpage 

We can also consider the effects of satellite contamination of the He-like features by modifying the theoretical $\mathcal{R} = f/i$ radial dependence (equation \ref{eq:Rofr}) by considering the ratio $\frac{f + f_{\rm sat}}{i + i_{\rm sat}}$. The \apec\/ model allows us to compute the sum $f + i$, which is not affected by the photoexcitation-driven alteration of the $f/i$ ratio, and also the ratios $\frac{f_{\rm sat}}{f+ i}$ and $\frac{i_{\rm sat}}{f+ i}$, where $f_{\rm sat}$ and $i_{\rm sat}$ represent the aggregate satellite line flux contributions to the $f$ and $i$ lines, respectively \citep{Porquet2001}. This requires a determination of each satellite's contribution to each of the $r$, $i$, and $f$ lines. We do this using the Gaussian-broadened model ($\sigma = 750$ \kms) to apportion a fraction of each satellite to each He-like component according to how much overlap there is in each profile. Quite a few of the satellites contribute significantly to both the $f$ and the $i$ lines (or the $i$ and the $r$ lines) due to the large wind Doppler-broadening value. 

\begin{figure}
\centering
\includegraphics[angle=0,width=0.40\textwidth]{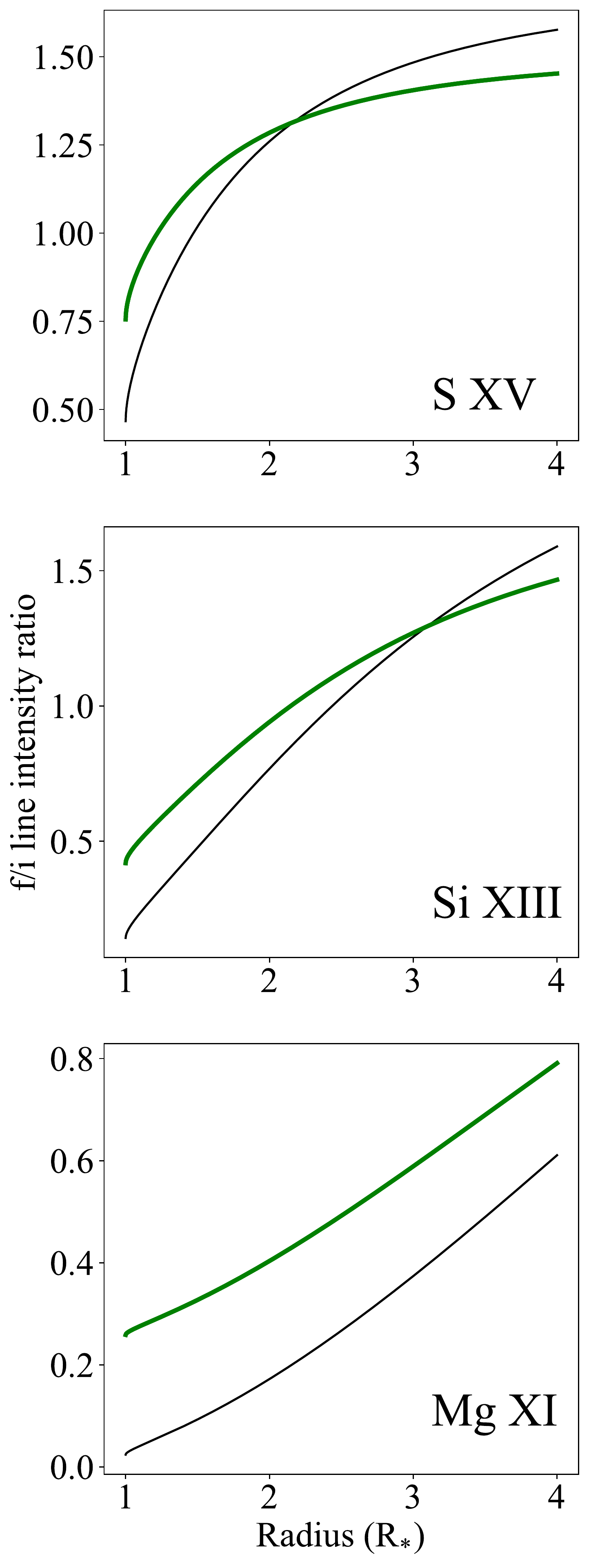}
    \caption{Adjusted theoretical models accounting for satellites (green) are compared to the models shown in Fig.\ \ref{fig:projRfir} (black). }
	\label{fig:model_comparison}
\end{figure}

In Fig.\ \ref{fig:model_comparison} we show the satellite-altered line ratio for each ion compared to the standard model as a function of radial location (and thus degree of UV photoexcitation). In each case the high-UV limit does not approach zero but rather approaches $\frac{f_{\rm sat}}{f+ i + i_{\rm sat}} > 0 $ and the low-UV limit approaches a value somewhat lower than $\mathcal{R}_{\rm o}$. These results are summarized in Table \ref{tab:satellites}. This latter trend is due to the fact that the low-UV limit of the $f/i$ is greater than unity while the satellite contribution to each of the two He-like components is similar. In principle, these altered models could be used to interpret measured $f/i$ ratios (from Gaussian profile fitting) that ignore the presence of satellites while fitting the data. Of course, the particular altered model curves we show here are specific to \zpup, according to its DEM and wind velocity, although these properties are relatively consistent among O star EWS sources \citep{Cohen2021}. In this paper, we use the standard models (black curves in Fig.\ \ref{fig:model_comparison} and Fig.\ \ref{fig:projRfir}) and account for the satellite lines directly in the data fitting.

\begin{table}
\caption{
Satellite contamination factors and altered $\mathcal{R} \equiv f/i$ in the asymptotic limits 
}
\centering

\begin{tabular}{ccccc}
  \hline
  ion & $\frac{f_{\rm sat}}{f + i}$ & $\frac{i_{\rm sat}}{f + i}$ & $\mathcal{R}_{\rm o, sat}$, $\mathcal{R}_{\rm o}$  & min $\mathcal{R}_{\rm sat}$ \\
  \hline
  S {\sc xv} & 0.463 & 0.352 & 1.52, 1.71 & 0.34 \\
    Si {\sc xiii} & 0.358 & 0.276 & 1.86, 2.38 & 0.28 \\
    Mg {\sc xi} & 0.286 & 0.221 & 2.16, 2.91 & 0.23 \\
    \hline
\end{tabular}
\\
{
The second and third columns contain the satellite line flux as a fraction of the total $f + i$ flux in our \apec\/ DEM model, assuming line profiles with 750 \kms\/ widths. The next column lists the no-UV-photoexcitation limit (and so the maximum possible $\mathcal{R}, \mathcal{R}_{\rm o, sat} \equiv (\frac{f + f_{\rm sat}}{i + i_{\rm sat}})_{\rm o}$), with the values ignoring satellites 
($\mathcal{R}_{\rm o} \equiv (\frac{f}{i})_{\rm o}$) listed also for comparison. The final column lists the minimum possible value of $\mathcal{R}$ in the presence of satellites, min $\mathcal{R}_{\rm sat} \equiv \frac{f_{\rm sat}}{f + i + i_{\rm sat}}$  . This latter value is zero in the absence of satellite line contamination. }
\label{tab:satellites}
\end{table} 

\vspace{0.3in}

\noindent

One interesting effect of the inclusion of the satellites in the spectral modeling is that oft-seen (inverse) trend of \Rfir\/ with atomic number in the context of Gaussian profile fitting \cite[e.g.][]{wc2007} is no longer present (though see the discussion of the mild \Ro\/ (inverse) trend with atomic number in \S5). This makes sense in the context of lower atomic number elements being in the low-forbidden line (low $\mathcal{R}$) limit where satellite contamination will have a greater impact on the relative flux near the forbidden line than it does for the higher atomic number elements.

\end{document}